\documentclass[useAMS,usenatbib]{mn2e}

\usepackage{graphicx}
\usepackage{float}
\usepackage{color}
\renewcommand{\P}{\mathcal{P}}
\newcommand{\M}{\mathcal{M}}
\newcommand{\D}{\mathcal{D}}
\renewcommand{\d}{\mathrm{d}}
\newcommand{\V}[1]{\mathrm{Var}\left[#1\right]}

\title[Gaussianisation for inference in cosmology]{Gaussianisation for fast and accurate inference from cosmological data}
\author[Robert L. Schuhmann, Benjamin Joachimi and Hiranya V. Peiris]{
Robert L. Schuhmann$^{1}$\thanks{E-mail:
\tt{robert.schuhmann.13@ucl.ac.uk}},
Benjamin Joachimi$^{1}$,
Hiranya V. Peiris$^{1}$\\
$^{1}$Dept. of Physics and Astronomy, University College London, Gower Place, London WCE1 6BT, UK\\}
\begin{document}

\date{Accepted --- Received ---; in original form ---}

\pagerange{\pageref{firstpage}--\pageref{lastpage}} \pubyear{2016}

\maketitle

\label{firstpage}

\begin{abstract}
We present a method to transform multivariate unimodal non-Gaussian posterior probability densities into approximately Gaussian ones via non-linear mappings, such as Box--Cox transformations and generalisations thereof. This permits an analytical reconstruction of the posterior from a point sample, like a Markov chain, and simplifies the subsequent joint analysis with other experiments. This way, a multivariate posterior density can be reported efficiently, by compressing the information contained in MCMC samples. Further, the model evidence integral (i.e. the marginal likelihood) can be computed analytically. This method is analogous to the search for normal parameters in the cosmic microwave background, but is more general. The search for the optimally Gaussianising transformation is performed computationally through a maximum-likelihood formalism; its quality can be judged by how well the credible regions of the posterior are reproduced. We demonstrate that our method outperforms kernel density estimates in this objective. Further, we select marginal posterior samples from {\it Planck} data with several distinct strongly non-Gaussian features, and verify the reproduction of the marginal contours. To demonstrate evidence computation, we Gaussianise the joint distribution of data from weak lensing and baryon acoustic oscillations (BAO), for different cosmological models, and find a preference for flat $\Lambda$CDM. Comparing to values computed with the Savage-Dickey density ratio, and Population Monte Carlo, we find good agreement of our method within the spread of the other two.
\end{abstract}


\section{Introduction}
According to the Bayesian paradigm, inference on any data set will yield a posterior probability distribution on the space of model parameters. This density function represents, in its entirety, the full knowledge gained in the attempt to infer the underlying parameters. Such distributions often depart significantly from a Gaussian form. This led to the widespread use of Monte Carlo sampling methods to report the typically non-Gaussian posterior constraints obtained from experiments, such as {\it Planck}\footnote{See \cite{Planck13, Planck15}. For the Markov chains see {\tt http://www.cosmos.esa.int/web/planck/pla}; consult {\tt http://lambda.gsfc.nasa.gov} for an eclectic list of data combinations in various cosmological models.}. Reconstructing the posterior density from such a Markov Chain Monte Carlo (MCMC) sample, e.g. to visualise the multivariate parameter constraints, or to combine the constraints from multiple data sets, can be nontrivial due to the large sample size necessary to appropriately map the distribution; in addition, the contours often need further smoothing for stylistic reasons.\\
Instead, we propose to redefine the underlying model parameters, so that the new posterior density approximately takes a Gaussian shape after the transformation from old to new parameters; this presupposes that we begin with a uni-modal posterior density. Such a transformation would allow for enormous data compression: instead of a full MCMC sample from the posterior distribution, we only need to report the Gaussianising transformation, and the first and second moments of the resulting Gaussian distribution. From these alone, we can reconstruct an analytic expression for the full non-Gaussian posterior density, and subsequently combine it with other data sets.\\ 
Further, it becomes possible to display and compare non-ellipsoidally-shaped contours of non-Gaussian parameter constraints -- whether joint or marginalised -- without any smoothing. Thus, this method allows for summarising posterior densities in a versatile and efficient way, which faithfully reproduces the information contained in the full probability density.\\
The idea of transforming a function to a Gaussian shape is, in principle, not limited to reproducing probability densities. As the integral over a Gaussian can be performed analytically, this opens up a strategy to feasibly compute high-dimensional integrals, such as the model evidence (i.e. the marginal likelihood).\\
The transformed model parameters are analogous to the normal parameters of the cosmic microwave background (CMB): it has been highly advantageous for rapid likelihood calculation (such as CMBfit, CMBwarp, and PICO, see \citealt{kosowsky, cmbfit, sandvik,cmbwarp, pico}), to redefine the cosmological model parameters such that the model is approximately linear in these newly defined {\it normal} parameters. Thus the likelihood approximately takes the form of a multivariate Gaussian density. For most observables, we would be at a loss to search for a linearising redefinition of the model parameter space directly motivated by the structure of the model itself. Instead, is it possible to computationally find suitable parameters, i.e. a suitable bijective transformation which approximately Gaussianises the posterior in question? \\
Extending the work of \cite{joachimi}, we present an algorithm to find and test such a non-linear Gaussianising transformation from a Markov chain sampling the posterior distribution of the original parameters. In principle, this distribution could stem from any experiment or data type. In Sect.~\ref{sec:gaussianisation}, we describe the details of the algorithm, verification of the reconstructed posterior distribution, and the specific transformations employed. Following an illustration of these on a toy example in Sect.~\ref{sec:toy},  Sect.~\ref{sec:performance} demonstrates the performance of our implementation, using Markov chains from the {\it Planck} satellite constraints on cosmological models (\citealp{Planck13, Planck15}). In Sect.~\ref{sec:evidence}, we present an efficient way to calculate the model evidence, a quantity needed to judge the predictivity of different model parameter spaces, via Gaussianising transformations. Section~\ref{sec:conclusio} offers conclusions and suggests future directions.

\section{Gaussianisation}
\label{sec:gaussianisation}
To find the right multivariate transformation, we will at first adopt the strategy of redefining each model parameter separately, i.e. the first new model parameter will only depend on the first old model parameter, etc. In Sect.~\ref{ssec:multipass}, we will drop this assumption and consider transformations which can correlate the model parameters.\\
The set of all multivariate Gaussianisation transformations, from which we are to pick the optimal one, will be constructed in the following way: assume a family of bijective real-valued functions $F_\Delta:\mathbfss{R}\rightarrow\mathbfss{R}$ indexed by  $n$ real transformation parameters $\Delta=(\delta^1, \ldots, \delta^n)$. Given the $d$-dimensional vector of model parameters $\bmath{X}=(X_1, \ldots, X_d)$, we transform to the new (Gaussian-distributed) parameters $\bmath{Y}$ via 
\begin{equation}
 \bmath{Y}=(Y_1, \ldots Y_d)=\left[F_{\Delta_1}(X_1), \ldots, F_{\Delta_d}(X_d)\right],\label{eq:trafo}
\end{equation}
where the full multivariate transformation is now specified by all $d$ transformation parameter $n$-tuples $(\Delta_1, \ldots, \Delta_d)$, i.e. one $\Delta_i=(\delta^1_i,\ldots,\delta^n_i)$ for each model parameter. To avoid confusion, we shall from now on distinguish between model parameters (MP), which the posterior probability density depends on, and transformation parameters (TP), which specify one Gaussianising transformation. The algorithm can be applied to arbitrary parametrised transformation families, suitable for various forms of non-Gaussianity -- in principle, we could even choose different transformations for each model parameter, instead of using the same shape $F_{\Delta_i}(X_i)$ for all of them.\\
Assuming such a bijective transformation $\bmath{X}\mapsto\bmath{Y}$, we immediately have an analytic form for the posterior density 
\begin{eqnarray}
\Pi(\bmath{X})&=&\left|\frac{\d\bmath{Y}}{\d\bmath{X}}\right|\tilde{\Pi}(\bmath{Y})\nonumber\\
&=&\left(\prod\limits_{i=1}^{d}\left|\frac{\d F_{\Delta_i}}{\d X}(X_i)\right|\right)\frac{1}{\sqrt{(2\pi)^d\det\bmath{\tilde\Sigma}}}\label{eq:ap}\\
&&\times\exp\left\{-\frac{1}{2}\left[\bmath{Y}(\bmath{X})-\tilde\bmu\right]^T\bmath{\tilde{\Sigma}}^{-1}\left[\bmath{Y}(\bmath{X}	)-\tilde\bmu\right]\right\}.\nonumber
\end{eqnarray}
One still needs to find the mean vector $\tilde{\bmu}$ and the covariance matrix $\tilde{\mathbf{\Sigma}}$ of the transformed posterior density $\tilde{\Pi}$. These are estimated from the transformed sample (see Sect.~\ref{ssec:optimize}).

\subsection{Finding the optimal transformation}
\label{ssec:optimize}
Given a weighted point sample $\mathcal{D}=\left\{(\bmath{X}^a, w^a)\right\}_{a=1}^\mathcal{N}$, containing $\mathcal{N}$ points in $\mathbfss{R}^d$ and probability weights $w^a$, which has been sampled from the posterior distribution in question, we wish to quantify the Gaussianisation properties of different transformations applied to this sample. To this end, we follow \cite{BoxCox} (see also \citealp{velilla} and \citealp{joachimi}) in maximising the profile likelihood over TP space, i.e. depending only on the $n\times d$ real transformation parameters contained in $\bmath{\Delta}=(\Delta_1,\ldots,\Delta_d)$. This likelihood is a function of the transformation parameters $\Delta$, quantifying how well each transformation Gaussianises the distribution of data set $\D$; however, it does not pertain to the posterior density in Eq.~(\ref{eq:ap}), which is a function of the model parameters $\bmath{X}$.\newline 
For the Gaussian parameters  $\tilde{\bmu}$, $\bmath{\tilde\Sigma}$ in Eq.~(\ref{eq:ap}), we insert their standard debiased weighted maximum-likelihood estimators, which depend on the transformed sample $\{(\bmath{Y}^a, w^a)\}_{a=1}^\mathcal{N}$ 
\begin{eqnarray}
\tilde{\bmu}&=&\frac{1}{W_1}\sum\limits_{a=1}^{\mathcal{N}}w^a\bmath{Y}^a;\\
\bmath{\tilde{\Sigma}}&=&\frac{W_1}{(W_1)^2-W_2}\sum\limits_{a=1}^{\mathcal{N}}w^a(\bmath{Y}^a-\tilde{\bmu})\,(\bmath{Y}^a-\tilde\bmu)^T,
\end{eqnarray}
with $W_1=\sum w^a$ and $W_2=\sum (w^a)^2$. These estimators depend on $\bmath{\Delta}$ indirectly, as they are computed after $\D$ has been transformed with $\bmath\Delta$. We arrive at the profile weighted log-likelihood
\begin{eqnarray}
\mathcal{L}(\bmath{\Delta}| \mathcal{D})&=&-\frac{W_1}{2}\ln\det\bmath{\tilde{\Sigma}}(\bmath{\Delta}, \D)\nonumber\\
&&+\sum\limits_{a=1}^{\mathcal{N}}w^a\sum\limits_{i=1}^d\ln\left|\frac{\d F_{\Delta_i}}{\d X}({X^a}_i)\right|,\label{eq:like} %change for weights!!!
\end{eqnarray}
where several terms independent of $\bmath\Delta$ have been discarded. In general, both the covariance matrix of the transformed sample and the Jacobian term will depend on the transformation parameters $\bmath{\Delta}$ in a non-linear way, hence finding the maximum-likelihood values for the TPs will require numerical optimisation. For this purpose, we have employed the GSL implementation of the well-known Nelder--Mead simplex algorithm (\citealt{NelderMead}).\\
As already noted by \cite{joachimi}, log-likelihood degeneracies in TP space are common. These may jeopardise the numerical stability of the calculation of $\mathcal{L}$. There are generic cases where a moderately large value for one transformation parameter may already result in unmanageably large numerical values for the transformed sample, such as e.g. the power transformation $X_i\mapsto(X_i)^{\lambda_i}$ with $\lambda_i\sim50$. Generically, the optimisation algorithm tends to slide into these TP space regions quite easily. Hence, we include a penalty term of the form
\begin{equation}
 \mathcal{P}(\bmath{\Delta})=\epsilon\sum\limits_{i=1}^d\sum\limits_{s=1}^n(\delta_i^s-\delta^{s,U})^p,\label{eq:penalty}
\end{equation}
where $\delta^{s,U}$ are the parameter values corresponding to the identity transformation. We minimise the function $-\mathcal{L}(\bmath{\Delta}; \mathcal{D})+\mathcal{P}(\bmath{\Delta}) $ over the $n\times d$ real numbers in $\bmath{\Delta}$. Values of $p=4$ and $\epsilon=10^{-4}$ have proven to be highly stabilising, and at the same time do not distort the shape of the resulting analytic posterior distribution.\\
In this work, we employ the Nelder--Mead algorithm just for illustrating the method -- faster and more reliable algorithms to find the global minimum of the likelihood function exist (such as Bound Optimisation BY Quadratic Approximation -- BOBYQA, see \citealp{bobyqa}) and can readily be applied here.

\subsection{Box--Cox transformations and their kin}
\label{ssec:boxcox}
The Box--Cox transformation (\citealp{BoxCox}) is a generalisation of the power map. This transformation family is widely used in statistics and econometrics, e.g. to make data approximately homoscedastic and normal. Our usage is different in that we use it to alter the distribution of model parameters, rather than the distribution of data. Including a shift parameter $a$, the one-dimensional version is defined as
\begin{equation}
   x\mapsto BC_{(a, \lambda)}(x)= \left\{
\begin{array}{ll}
      \lambda^{-1}[(x+a)^\lambda-1] & (\lambda\neq 0)\\
      \log(x+a) & (\lambda=0)\\
\end{array}\right.
\end{equation}
for a single MP $x$, i.e. $(\delta^1, \delta^2)=(a, \lambda)$. Note that the family is continuous at $\lambda=0$ and that the mapping requires $a<x$. %Generalizations?
Typically, an MP with a skewed distribution can be transformed to an MP with symmetric, Gaussian distribution upon the appropriate choice of the power TP $\lambda$, e.g., a log-normal distribution can be analytically transformed to a Gaussian with $a=\lambda=0$. The identity transformation corresponds to $\delta^{1, U}=a=1$ and $\delta^{2, U}=\lambda=1$. Inserting this transformation family into Eq.~(\ref{eq:ap}), we recover the formula given in \cite{joachimi}. \\
As an extension of the Box--Cox family, we propose the Arcsinh--Box--Cox transformation (`ABC transformation' hereafter):
\begin{equation}
   x\mapsto ABC_{(a, \lambda, t)}(x)= \left\{
 \begin{array}{ll}
      t^{-1}\sinh[t\,BC_{(a,\lambda)}(x)] & (t > 0)\\
      BC_{(a, \lambda)}(x) & (t=0)\\
      t^{-1}\,\mathrm{arcsinh}[t\,BC_{(a,\lambda)}(x)] & (t < 0).\\
 \end{array}\right.
\end{equation}
The inclusion of the TP $t$ will prove particularly useful to remove residual kurtosis from a MP distribution. The identity transformation reads $\delta^{1,U}=a=1$, $\delta^{2, U}=\lambda=1$, $\delta^{3,U}=t=0$.\\
The Box--Cox family does not form a group, because two subsequent transformations cannot be expressed as another Box--Cox transformation; the same holds for the ABC family. This will be of importance for Sect.~\ref{ssec:multipass}.\\
Box--Cox transformations demonstrate that the domain of the function $F_\Delta$ -- in particular its dependence on $\Delta$ -- requires special attention: for given $a$, it is defined only for $x\in(-a, \infty)$, the same holds for ABC transformations. Thus, the optimisation procedure for the sample $\D=\{\bmath{X}^a\}_{a=1}^\mathcal{N}$ requires that $a_i$, the shift parameter for the model parameter $X_i$, is bounded from below, i.e. $a_i>\min_a(-X^a_i)$. Conversely, this means that, once the optimal transformation parameters $\bmath{\Delta}^\mathrm{opt}$ are found and inserted into the analytic expression for the original posterior density, Eq.~(\ref{eq:ap}), it is not defined for every value possible value of the MP $\bmath{X}$, but only for $X_i>a_i^\mathrm{opt}$. This also necessitates that the normalisation needs to be adjusted, which can be done analytically. However, if the sample is large enough so that the tails of the distributions are properly represented, this truncation of the domain is not problematic.

\subsection{Verifying the optimal transformation}
\label{ssec:verify}
Once the optimal transformation within its family is found, how do we judge the effectiveness of the resulting Gaussianisation? We adopt the following pragmatic standpoint: if the analytic posterior manages to reproduce the one-dimensional and two-dimensional marginalised contours of the sample, it is deemed acceptable. To this end, we propose the test via a cross-contour (CC) plot. The idea is to characterise a probability density by the location of its contours - the surfaces of constant density - and the probability mass stored inside, i.e. the integral of the density over the interiour of a contour. If two densities $p(\bmath X)$ and $q(\bmath X)$ are identical, then they will store the same mass in any region of the parameter space; if they are different, we expect to find different probabilities for the same regions (e.g. the regions bounded by contours of $p$). Thus, looking at the family of contour-bounded regions of $p$, we can ask: does the probability for these, assigned via $q$, agree with the probability for them assigned via $p$?\\
To formalise this, consider the following: given a probability density $p$ in $d$ dimensions, which takes function values between 0 and $p_\mathrm{max}$, we define the contour-bounded region assigned to the density value $r\in[0, p_\mathrm{max}]$ as
\[
\Omega_p(r)=\{\bmath{X}\in\mathbfss{R}^d: p(\bmath{X})\geq r\}.
\]
The probability mass enclosed in any of these is
\[
\int_{\Omega_p(r)}p(\bmath{X})\,\d\bmath{X}\in[0,1].
\]
Now, assuming we have two probability densities $p$ and $q$ in $d$ dimensions, do the contours of $q$ reproduce those of $p$? They do in the relevant sense if for every $r\in[0,p_\mathrm{max}]$, the $q$-mass enclosed in the $r$-contour of $p$ equals the $p$-mass in this contour, i.e.
\begin{equation}
 \int_{\Omega_p(r)}q(\bmath{X})\,\d\bmath{X}= \int_{\Omega_p(r)}p(\bmath{X})\,\d\bmath{X}.\label{eq:mass}
\end{equation}
It should be noted that this alone is not a sufficient condition for $p\equiv q$, but the counterexamples, which can be constructed mathematically, are non-generic and can be neglected for our purposes.\\
To detect deviations of the contours of $p$ and $q$, we could simply plot the left and the right side of Eq.~(\ref{eq:mass}) for a grid of $r$-values between 0 and $p_\mathrm{max}$, and plot the points with respect to the line $y=x$. For concrete problems, it is often more instructive to subtract the right side from the left side, and plot the excess (or deficit) probability mass of $q$ inside the contours of $p$. If, in this plot, the excess for every contour is consistent with zero, we have succeded.\\
In our situation, we compare a point sample $\mathcal{D}$ with a probability density function $p$ -- the analytic posterior density as reconstructed via Gaussianisation. The right side of Eq.~(\ref{eq:mass}) is the probability mass in the region where the density is greater or equal to $r$; the left side is the fraction of the point sample which lies in the same region. So, for every value $r$ in the range of $p$, we find the probability mass enclosed in $\Omega_p(r)$ by gridding $p(\bmath{X})$ over a region containing the sample. Similarly, we count the number of points in $\mathcal{D}$ where the value of $p$ is above $r$, to compute the fraction of points that lie within $\Omega_p(r)$. This fraction is an estimator of the actual probability mass enclosed, because $\mathcal{D}$ is a discrete sample from the actual posterior distribution. To find the variance of this estimator, we calculate the fraction on 2,000 bootstrap realisations of $\mathcal{D}$, and determine the 95\%-confidence intervals from these. If, for every $r$, the analytic posterior probability mass inside $\Omega_p(r)$ is within this confidence interval for the sample point fraction within $\Omega_p(r)$, we judge our reconstruction attempt to be successful. \\
It should be noted that poor MCMC sampling of the original target density will yield a poor representation of this density by our reconstructed density (\ref{eq:ap}). Any information about the distribution lost by undersampling cannot be regained. However, as demonstrated in Sect.~\ref{sec:toy}, our method reproduces less biased contours than other standard methods of density estimation even in regions of low point density, i.e. where any density estimate must be an extrapolation. Hence, it can be used when the length of the input Markov chains is restricted by computational cost or file size.

\subsection{Multi-pass transformations}
\label{ssec:multipass}
If even the optimal Gaussianising transformation amongst a given family does not bring the posterior density sufficiently close to a Gaussian shape (e.g., as determined via a CC plot), we have two options. We can provide a different family of transformations and redo the optimisation; or we can repeat the process on the sample after the first transformation. As already mentioned in Sect.~\ref{ssec:boxcox}, the transformation families employed in this work do not form groups. Hence, two subsequent transformations do not result in another transformation from that family, and transforming twice potentially provides a better Gaussianisation than transforming once. In principle, it is possible to apply multiple subsequent transformations, should the quality of the result necessitate it.\\
In this spirit, we have implemented the following two-pass transformation protocol:
\begin{enumerate}
 \renewcommand{\theenumi}{Step \arabic{enumi}:}
 \item Optimise the TPs of the first transformation.
 \item Linear reshaping: centring, rescaling, rotating.
 \item Optimise the TPs of the second transformation.
\end{enumerate}
Strictly speaking, this transformation, whilst being bijective, no longer falls into the class as set up in Eq.~(\ref{eq:trafo}), as different model parameters are mixed. Nonetheless, Eq.~(\ref{eq:ap}) for the analytic posterior density generalises in a straightforward way.\\
In the second step, the sample after the first Gaussianising transformation is subjected to the following maps (in this order): subtract the sample mean from every parameter, so that the sample is centred on the origin. Then, rescale every parameter such that the standard deviation is unity. Finally, rotate into the eigenbasis of the covariance matrix -- this procedure is generally known as principal component analysis (PCA). These reshaping operations not only help to avoid numerical instabilities (centring, rescaling), but also open up new directions for Gaussianisation by presenting uncorrelated parameters to the second Gaussianising transformation, since the transformations defined in Eq.~(\ref{eq:trafo}) cannot mix parameters. If two parameters have substantial covariance after step 1, it can be crucial to decorrelate them.\\
Nevertheless, a price is to be paid for the Gaussianising power added with Step 2: it sacrifices a decisive property of the simple one-step transformation routine, namely that every transformed MP $Y_i$ only depends on a single untransformed MP $X_i$. This property allows for easy marginalisation of the analytic posterior: to compute this, we can marginalise the Gaussianised sample by dropping all coordinates we wish to marginalise out and determining the mean vector and covariance matrix of the remaining ones. Transforming this marginalised Gaussian density back will then yield the marginalised posterior density on the untransformed MPs. However, with linear reshaping included, this is no longer possible.\\
This may be problematic for some applications (such as visualisation of 1D or 2D marginal distributions, or creating a CC plot), but not for others -- as long as we need only the marginal distribution of a single combination of parameters, we can marginalise by discarding all MP columns of the sample except the ones in question, prior to Gaussianising.
\begin{figure}
  \centering
  \begin{minipage}{.5\textwidth}
   \centering
   \includegraphics[width=\linewidth]{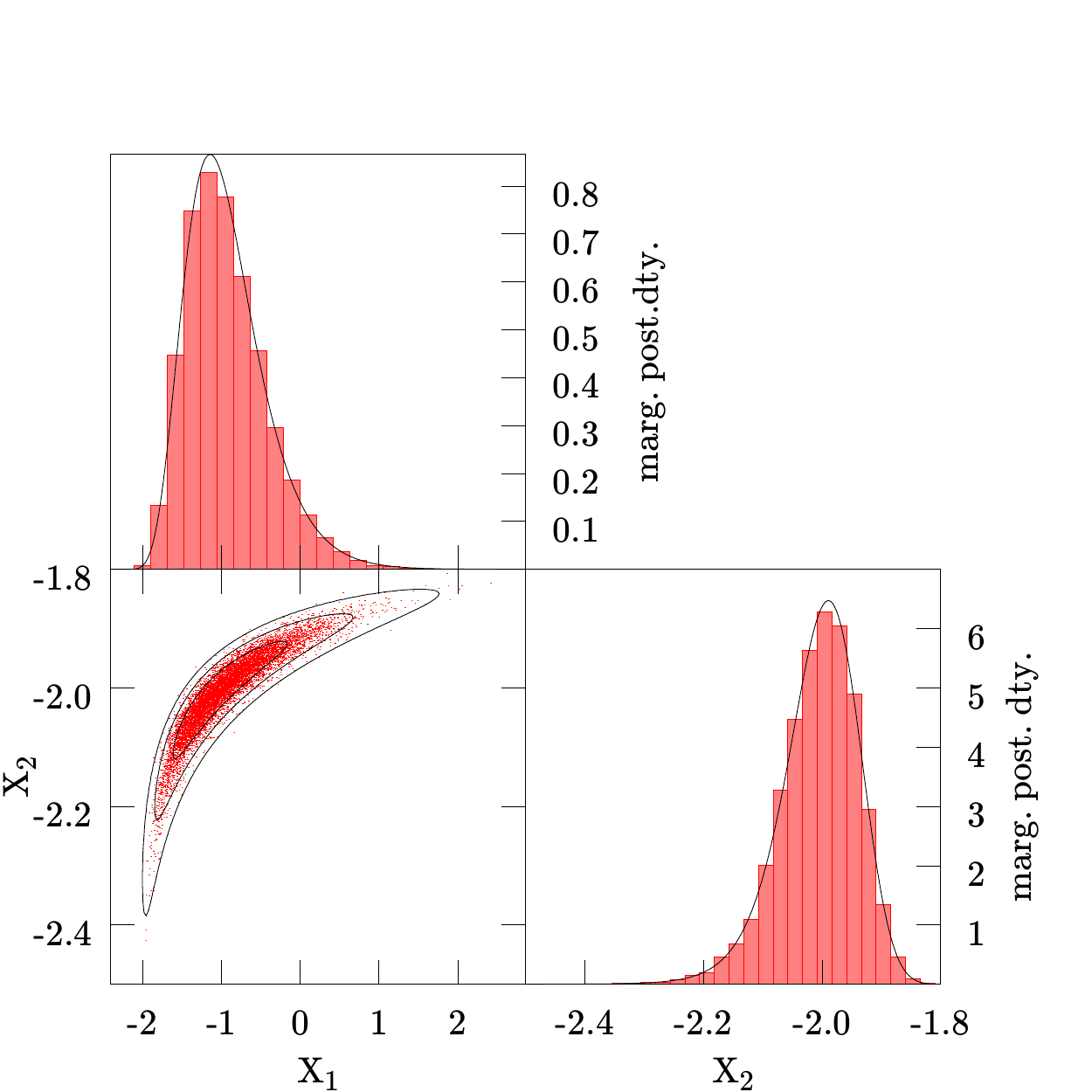}
  \end{minipage}
  \begin{minipage}{.5\textwidth}
   \centering
   \includegraphics[width=\linewidth]{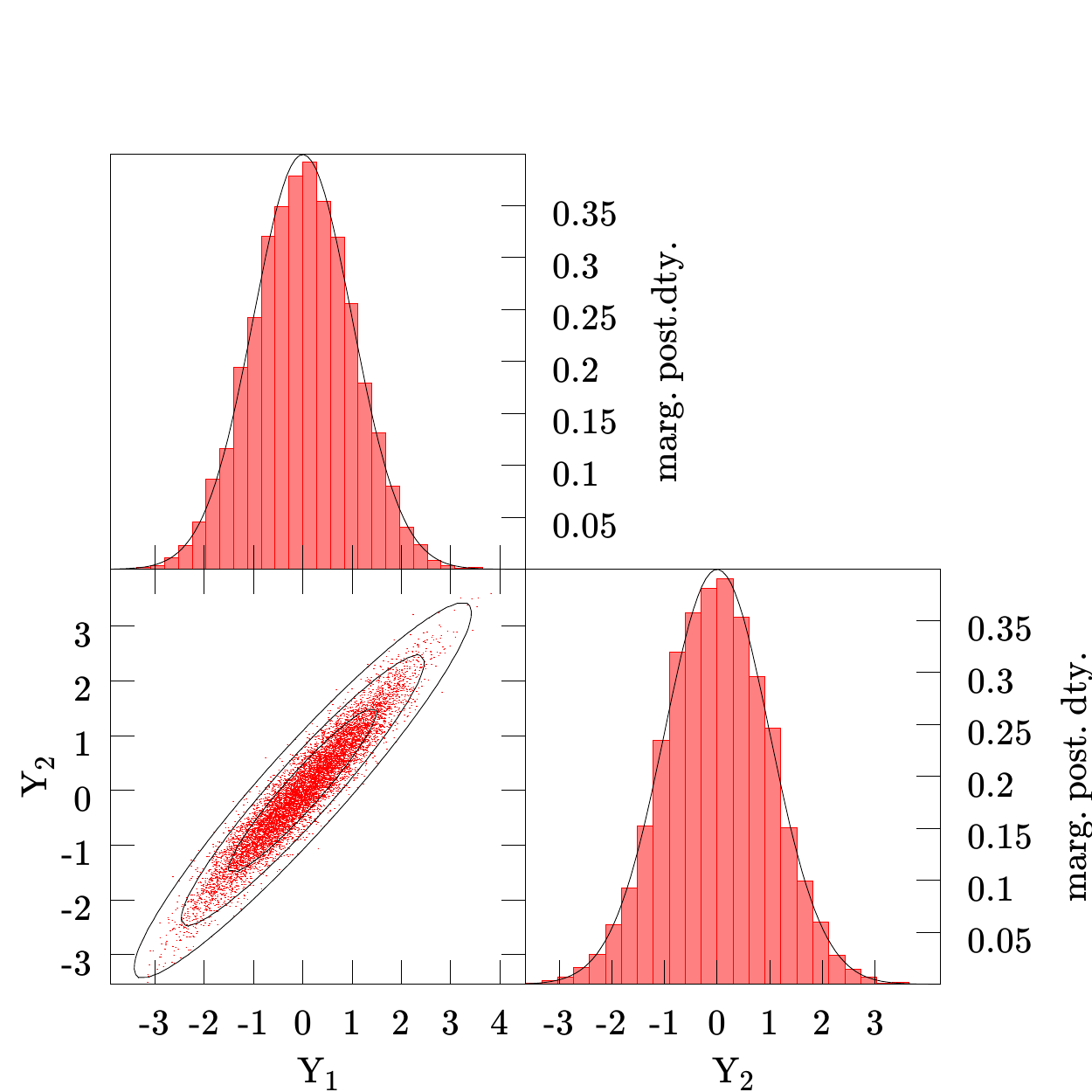}
  \end{minipage}
  \caption{Bivariate sample before (top) and after Gaussianisation (bottom). We show the two-dimensional sample and its 1D marginals (red), and compare to the reconstructed analytic posterior density (black): the full 2D contours, and its 1D marginal distributions.}
  \label{fig:toy_multivar}
\end{figure}
\section{A toy example}
\label{sec:toy}
We illustrate these ideas on a two-dimensional example. We draw a sample of 10,000 points from a bivariate Gaussian distribution, and map it through an inverse Box--Cox transformation with known input TP values (see Table~\ref{tab:params}). All weights are set to unity. \\
This mock data sample has the advantage that there is at least one Box--Cox transformation which precisely Gaussianises the underlying probability distribution. Figure~\ref{fig:toy_multivar} shows the original sample, and the one transformed with the one-pass Box--Cox transformation which was found to be optimally Gaussianising, i.e. maximising the log-likelihood in Eq.~(\ref{eq:like}). As this is a comparably simple problem, we have set the penalty term in Eq.~(\ref{eq:penalty}) to zero. The Nelder--Mead algorithm was started sixteen times independently with randomised initial conditions. The values of the recovered optimal TPs are shown in Table~\ref{tab:params}; the standard deviation amongst these sixteen values is of order $10^{-7}$ at worst, so multiple Nelder--Mead runs are not necessary in this low-dimensional example: all of them find the same maximum of the log-likelihood. In high-dimensional cases, however, this strategy can increase the robustness of the procedure.
\begin{figure}
  \centering
  \begin{minipage}{.49\textwidth}
   \centering
   \includegraphics[width=\linewidth]{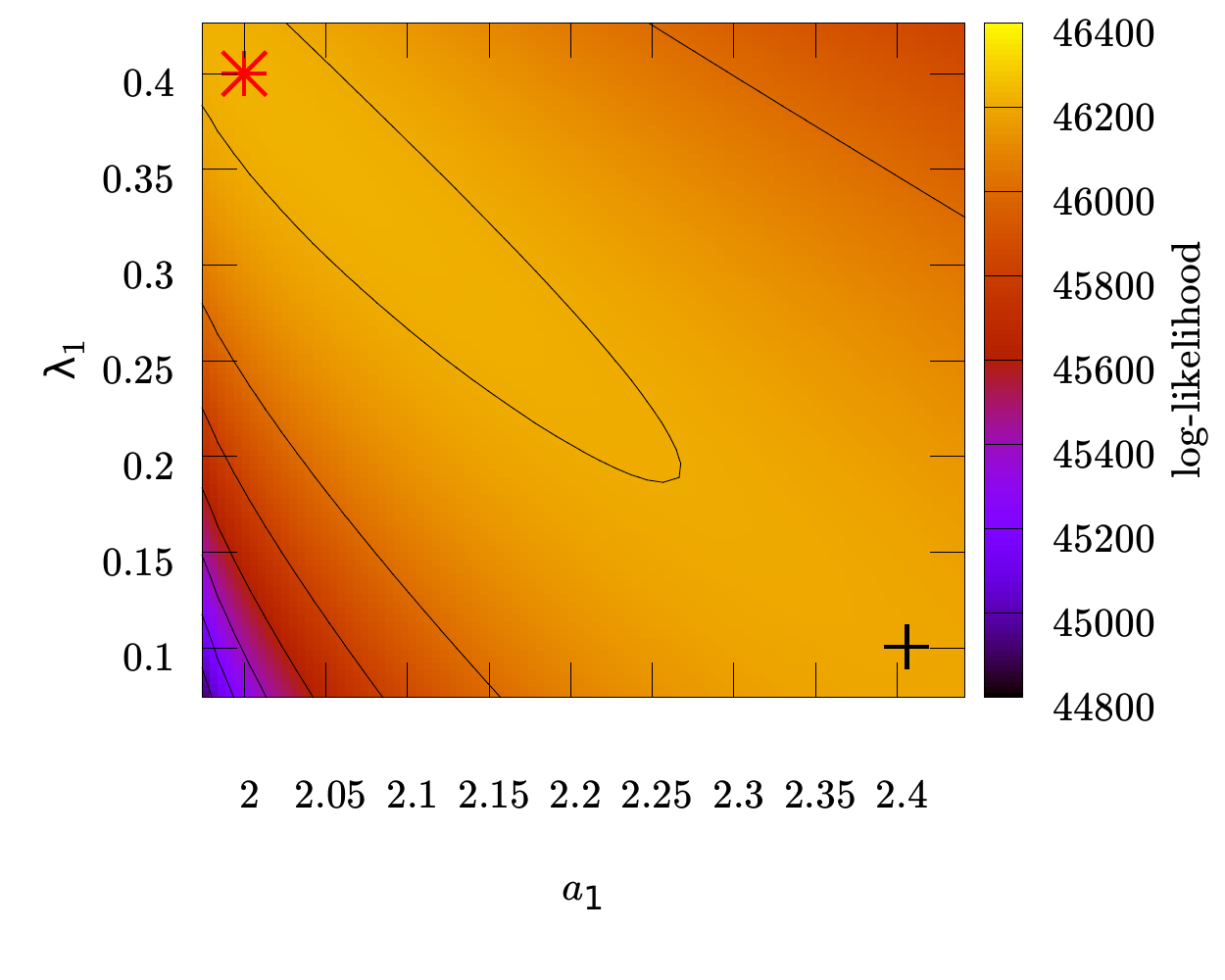}
  \end{minipage}
  \begin{minipage}{.49\textwidth}
   \centering
   \includegraphics[width=\linewidth]{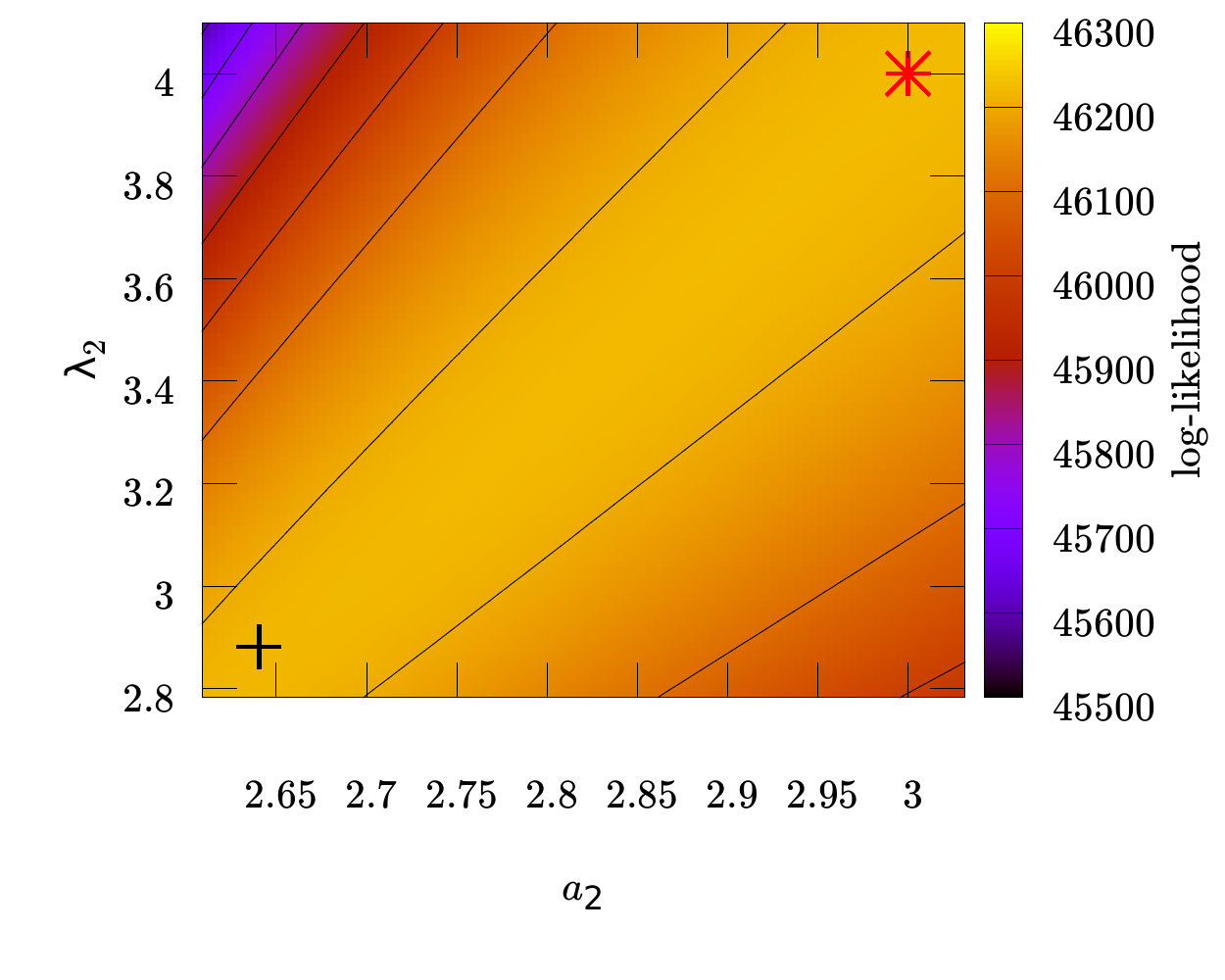}
  \end{minipage}
  \caption{Profile log-likelihood for the transformation parameters relating to $X_1$ (top), and to $X_2$ (bottom). The red star shows the input values, the black cross the recovered values, as projected onto the plane. The degeneracies between different transformation parameters are apparent.}
  \label{fig:toy_likelihood}
\end{figure}
The apparent difference between the parameters of the single inverse Box--Cox transformation and the values found for Box--Cox optimisation is due to degeneracies in parameter space. To illustrate these, we show the profile likelihood for $(a_1, \lambda_1)$ where $(a_2, \lambda_2)$ are held fixed at their input values, and vice versa, in Fig.~\ref{fig:toy_likelihood}. The TPs found by the optimisation algorithm (black crosses - note that they are projected onto the plane for which the profile likelihood is shown) are degenerate with the input ones (red star). Both Box--Cox transformations map the distribution to sufficiently Gaussian form.
\begin{figure}
 \centering
 \begin{minipage}{.49\textwidth}
  \centering
  \includegraphics[width=\linewidth]{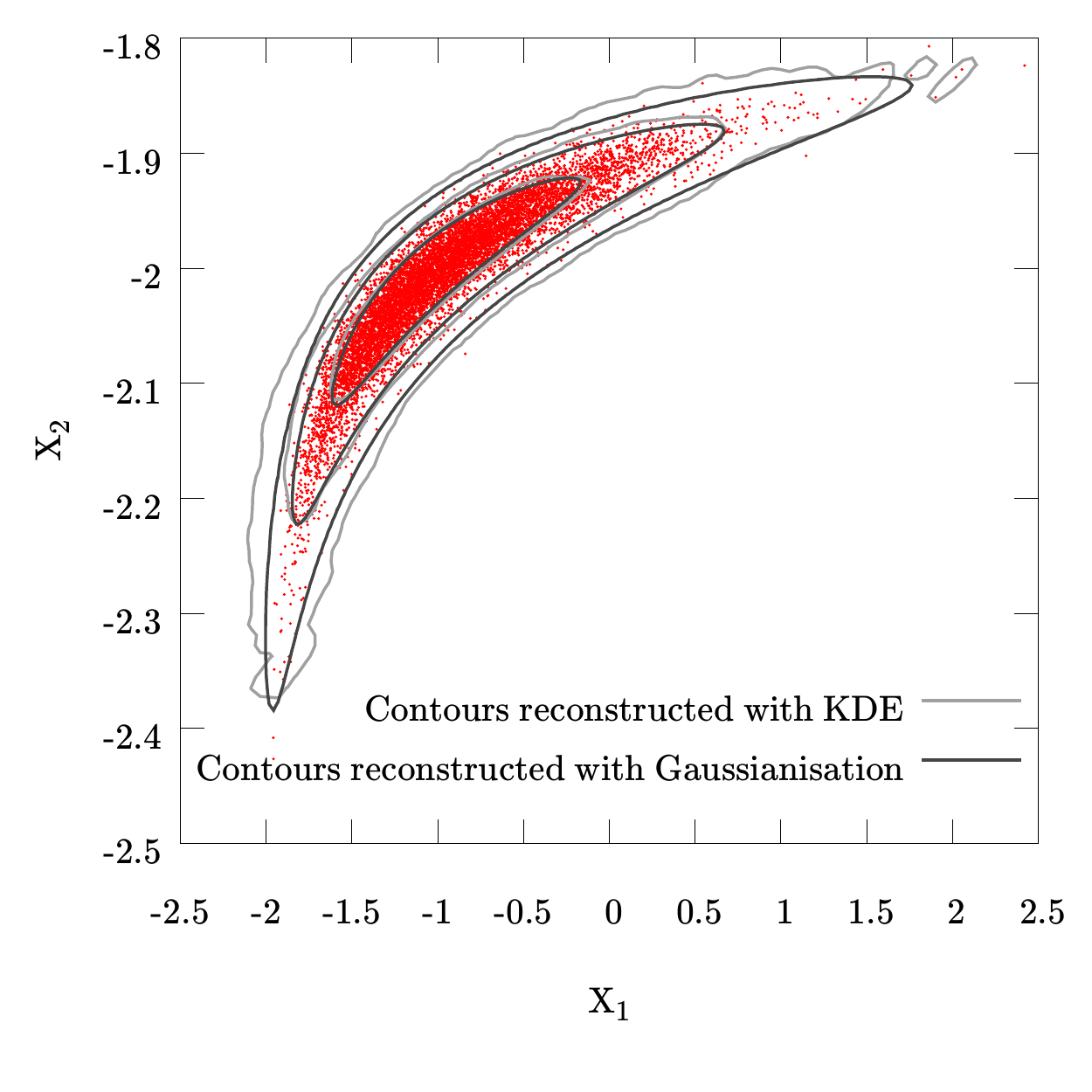}
 \end{minipage}
 \begin{minipage}{.49\textwidth}
  \centering
  \includegraphics[width=\linewidth]{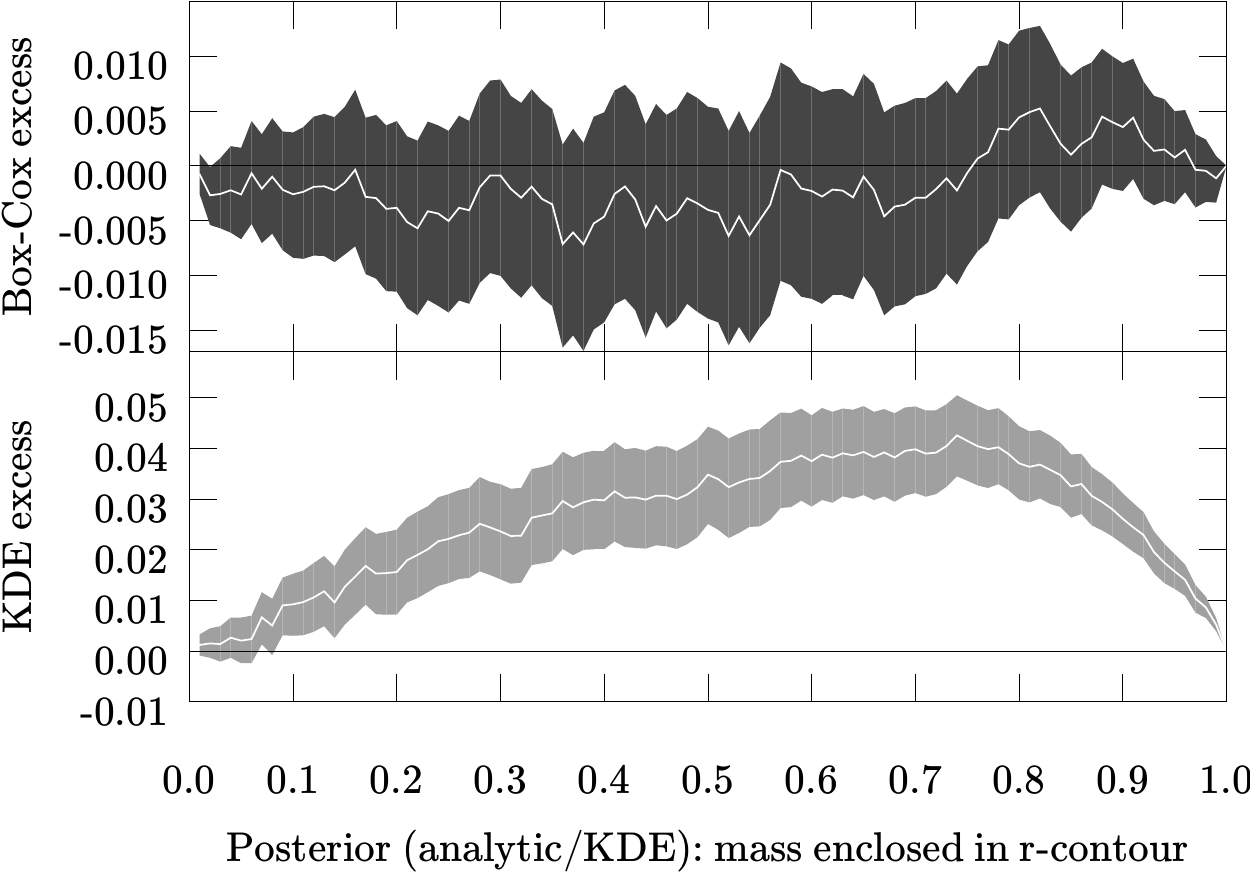}
 \end{minipage}
 \caption{Comparison of the analytic posterior density, as found via Gaussianisation (dark gray), to kernel density estimation (light gray). The  top panel shows the 2D contours of each density estimation method in relation to the original sample (red dots). The bottom panel (CC plots; top: Box--Cox, bottom: KDE) compares the respective contours of each function to the original sample: We determine the fraction of the point sample located inside one probability contour, and plot the excess of this fraction over the probability density mass for that same contour. The band shows the 95\%-variance in the point fraction due to sampling.}
  \label{fig:toy_ap_vs_kde}
\end{figure}
\begin{table}
\centering
\begin{tabular}{c|c|c}
\hline
Parameter & input value & recovered value\\
\hline
$a_1$ & 2 & $2.4$\\
$\lambda_1$ & 0.4 & $0.1$\\
$a_2$ & 3 & $2.6$\\
$\lambda_2$ & 4 & $2.9$\\
\end{tabular}
\caption{Optimally Gaussianising parameters for the distribution in Fig.~\ref{fig:toy_multivar}, as found with one-pass Box--Cox transformation.}
\label{tab:params}
\end{table}
We compare our method of reconstructing an analytic posterior density from an MCMC sample with the standard nonparametric method, kernel density estimation (KDE), which also aims to find a functional form for the probability density. The $1-3\sigma$-contours of the posterior density from Gaussianisation are shown jointly with those from KDE: these employ a Gaussian kernel, whose covariance matrix is estimated from the sample, and Silverman's rule (\citealt{silverman}) has been used to determine the bandwidth parameter. No additional smoothing has been applied in Fig.~\ref{fig:toy_ap_vs_kde}, top panel. The bottom panel shows the excess cross-contour probability masses between analytic posterior and sample, and KDE and sample respectively, as detailed in Sect.~\ref{ssec:verify}. Whereas the Box--Cox posterior is consistent with the sample distribution for every single contour, the KDE contours show a strong bias -- the contours are wider than they should be. Given that the precision of the contour reconstruction is, for the Box--Cox method, limited only by the finite size of the sample, it has the potential to perform better than the (biased) kernel density method.
Additionally, for applications in which frequent calls of the posterior density are a bottleneck for computation speed, our method of density reconstruction can be advantageous: the additional initial cost for finding the transformation parameters can be outweighed by the subsequent evaluation speedup.

\section{Performance results: \textbfit{P\lowercase{lanck}} data}
\label{sec:performance}
\begin{figure*}
  \centering
  \begin{minipage}{.49\textwidth}
   \centering
   \includegraphics[width=\linewidth]{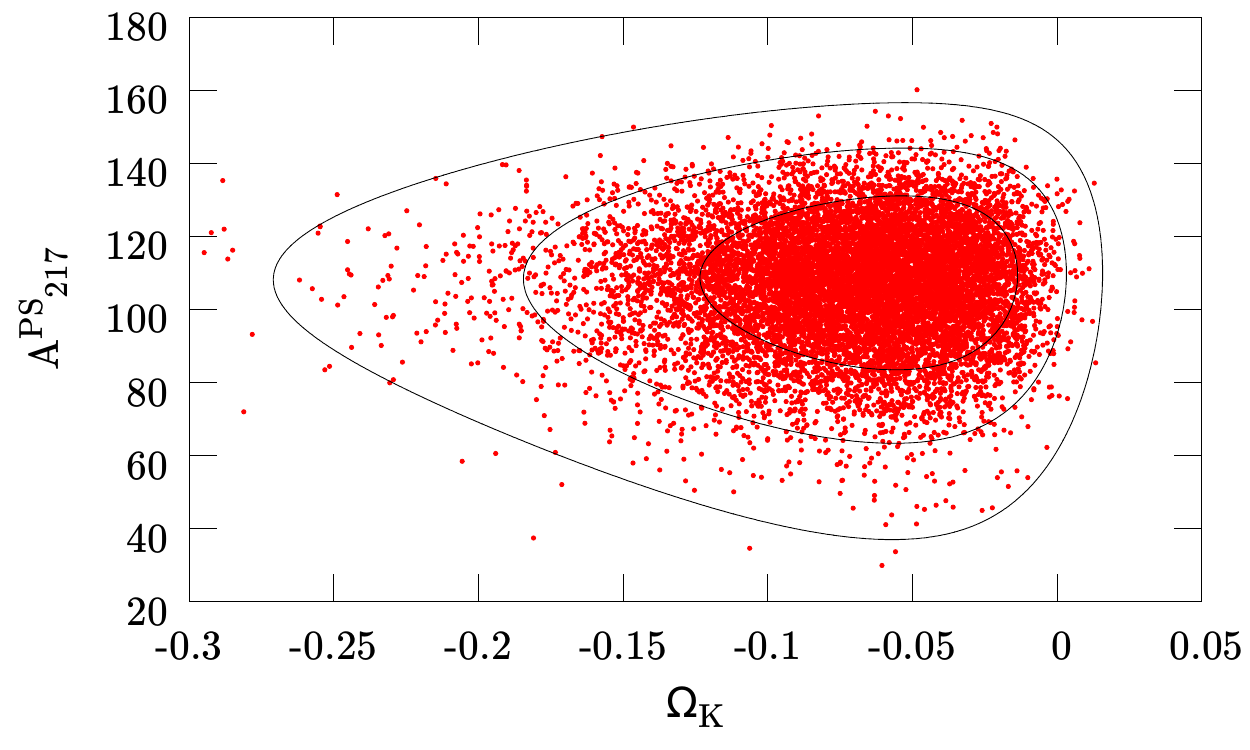}
  \end{minipage}
  \begin{minipage}{.49\textwidth}
   \centering
   \includegraphics[width=\linewidth]{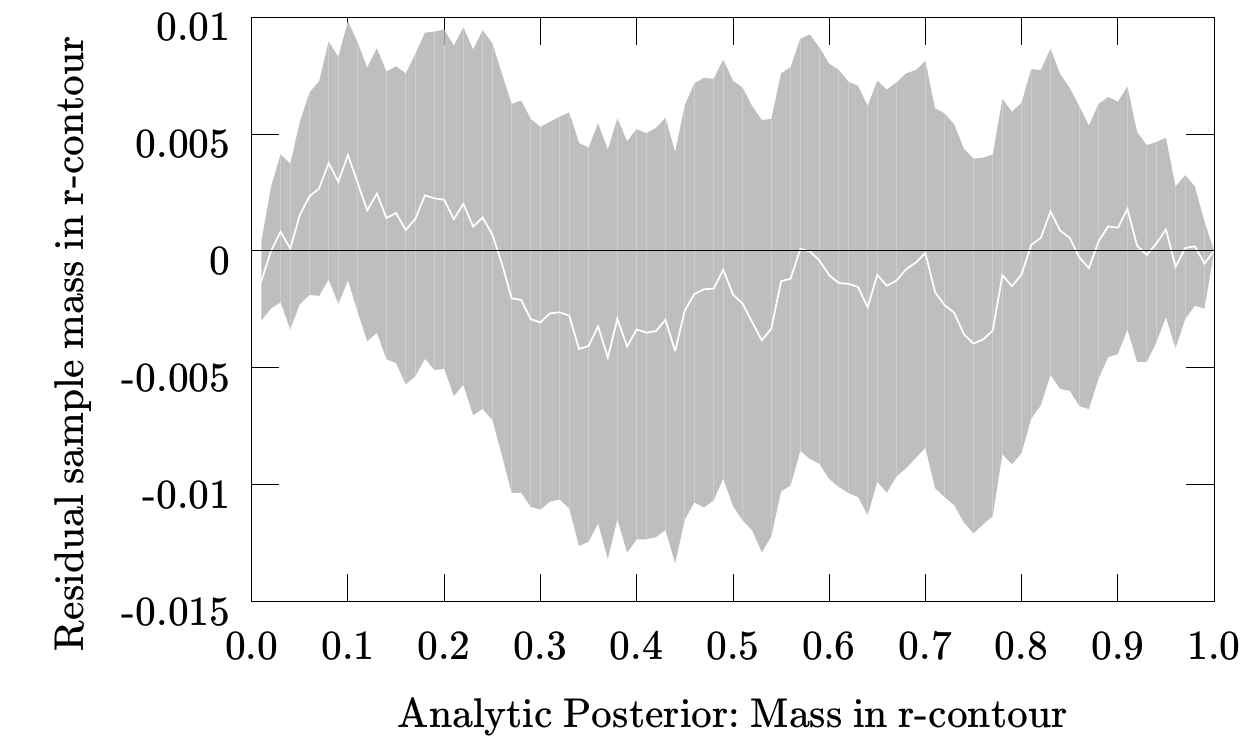}
  \end{minipage}
  \caption{One-pass Gaussianisation of a triangular-shaped non-Gaussian feature in a 2D marginal {\it Planck} posterior via ABC transformation. {\it Left}: original sample (red dots) and reconstructed analytic posterior (black contours). {\it Right}: the CC plot shows that for every contour of the analytic posterior, the probability mass inside (white line) equals the fraction of the point sample inside, within its 95\%-confidence interval (green band).}
  \label{fig:triangle_onepass}
\end{figure*}
\begin{figure*}
 \centering
 \includegraphics[width=0.98\textwidth]{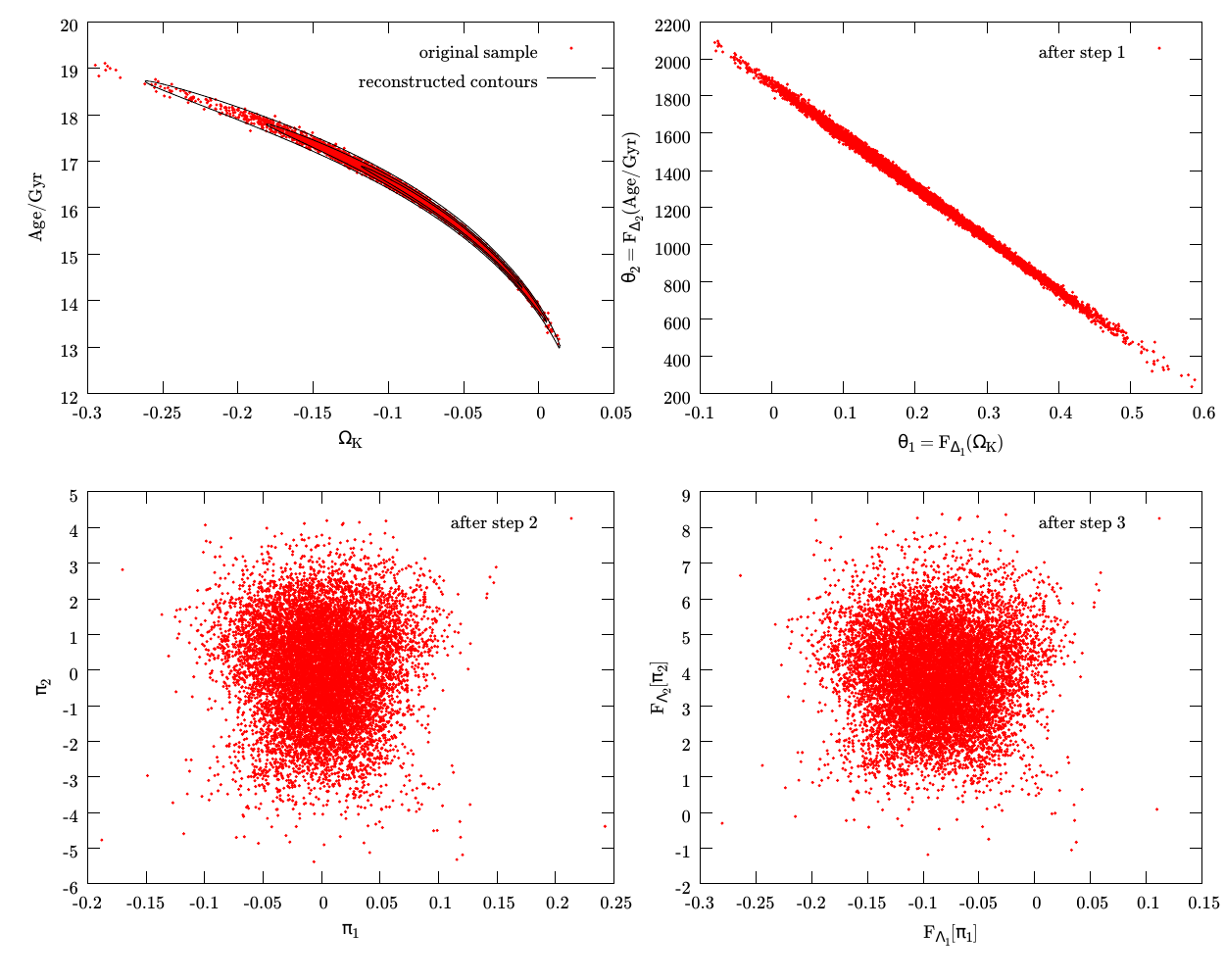}
 \caption{Two-pass Gaussianisation of a non-Gaussian model parameter degeneracy in a 2D marginal {\it Planck} posterior via ABC transformation, explicitly showing the protocol described in Sect.~\ref{ssec:multipass}: $(\theta_1, \theta_2)$ are the parameters after the first transformation; $(\pi_1, \pi_2)$ are the coordinates after rotation into the PCA eigenbasis of the centred and rescaled $(\theta_1, \theta_2)$-sample, which are finally transformed again. $(\Delta_1, \Delta_2)$ designate the transformation parameters of the first, $(\Lambda_1, \Lambda_2)$ those of the second ABC transformation. Note how crucial the intermediate PCA step is to achieve Gaussianity.}
 \label{fig:banana_twopass_protocol}
\end{figure*}
\begin{figure*}
 \centering
  \includegraphics[width=\textwidth]{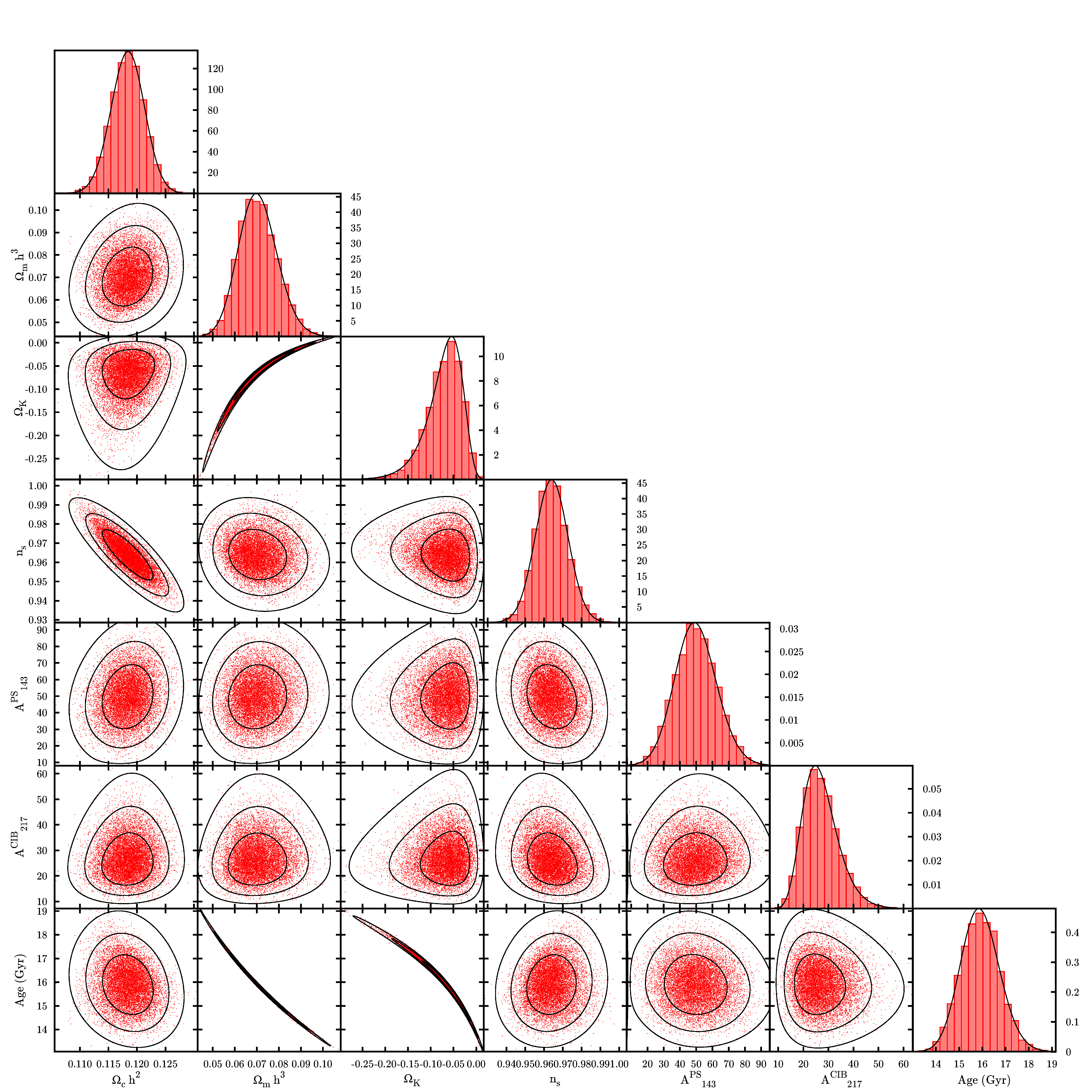}
 \caption{Reconstruction of a seven-dimensional {\it Planck} posterior density via a one-pass ABC transformation: 1D and 2D marginals. Black: marginal analytic posterior density (1D) or 1,2,3 $\sigma$ contours. Red: marginal point sample distributions. For the 1D cases, the histograms have renormalised bar heights, to demonstrate the agreement with the value of the probability density.}
 \label{fig:stairs}
\end{figure*}

To demonstrate how the algorithm works on real data, we have employed MCMC samples from the first data release of the {\it Planck} mission (see \citealp{Planck13}). This satellite has measured the temperature and polarisation anisotropies in the CMB, whose power spectra are sensitive measures of the underlying cosmology. The {\it Planck} Collaboration has published several data products\footnote{See {\tt http://www.cosmos.esa.int/web/planck/pla}.}, including MCMC samples from the posterior probability densities of various cosmological models, generated with CosmoMC (see \citealp{cosmomc}, also: {\tt http://cosmologist.info/cosmomc}). \\
The baseline cosmology is the standard model of a flat Universe with cold dark matter and a cosmological constant, commonly known as $\Lambda$CDM. It contains six parameters: $\Omega_bh^2$ (today's baryon density), $\Omega_ch^2$ (today's cold dark matter density), $100\,\theta_\mathrm{MC}$ (scaled sound horizon), $\tau$ (reionisation optical depth), $n_s$ (spectral index of primordial scalar perturbations), and  $\ln(10^{10}A_s)$ (log power amplitude of primordial scalar perturbations). Several extensions of this baseline model are also listed, including those by adding either of the following parameters: $\Omega_K$ (curvature parameter), $w$ (dark energy equation of state), $r$ (primordial tensor-to-scalar amplitude ratio), and $N_\mathrm{eff}$ (effective number of relativistic degrees of freedom). Further, these chains list derived quantities, e.g. today's Hubble parameter $H_0$, the age of the Universe, and a variety of foreground modelling parameters, such as $A^\mathrm{PS}_\nu$ and $A^\mathrm{CIB}_\nu$, modelling the amplitudes of Poisson point sources and the cosmic infrared background in the frequency bands $\nu=100\,\mathrm{GHz}$, $143\,\mathrm{GHz}$ and $217\,\mathrm{GHz}$. These are of particular interest to us, as the most prominent non-Gaussian features of the posterior densities can be seen in them.\\
\begin{figure*}
 \centering
 \includegraphics[width=0.98\textwidth]{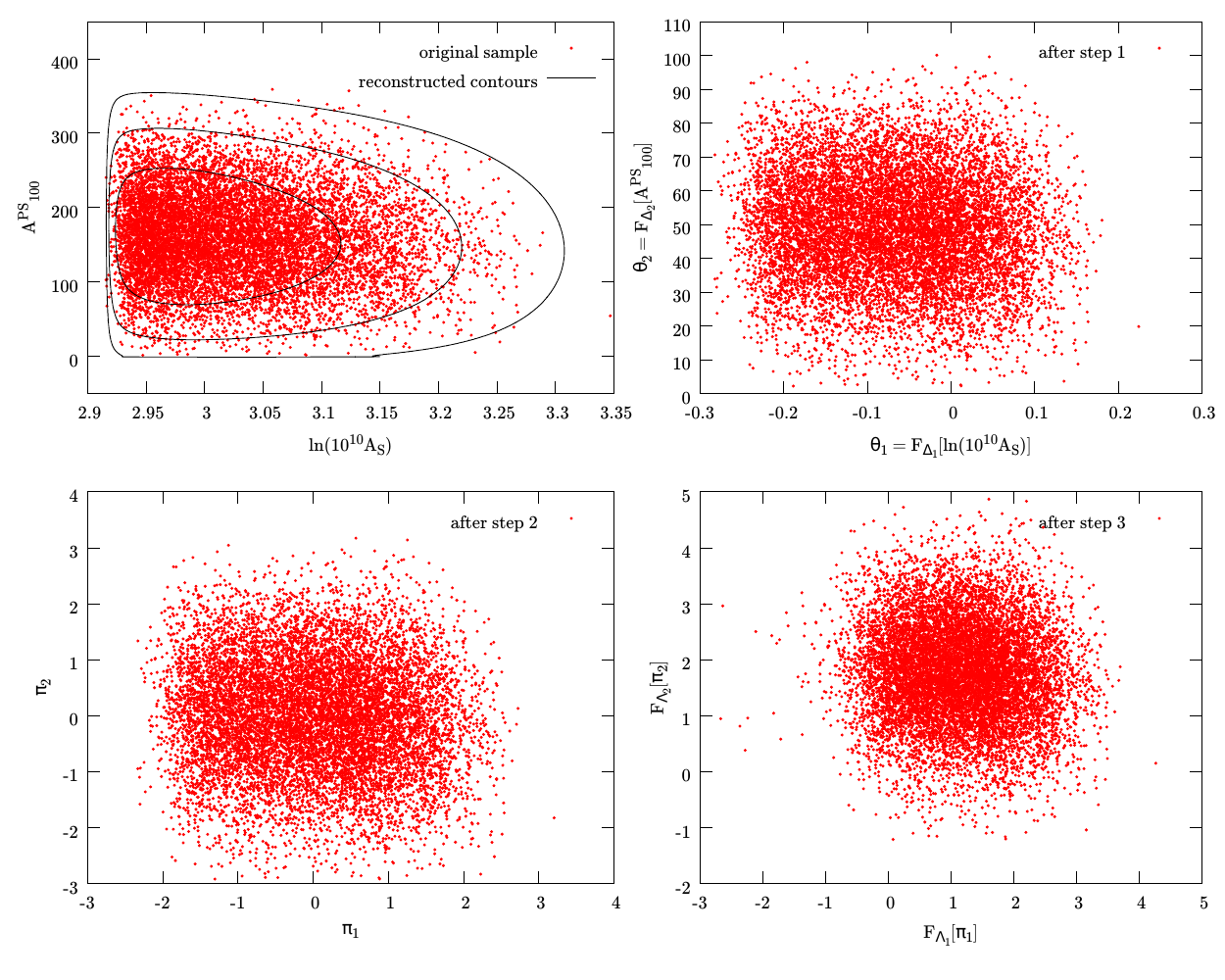}
 \caption{Two-pass Gaussianisation of a wall-like non-Gaussian feature in a 2D marginal {\it Planck} posterior via ABC transformation; in the same format as as Fig.~\ref{fig:banana_twopass_protocol}, but with the transformation parameters $(\Delta_1, \Delta_2)$ and $(\Lambda_1, \Lambda_2)$ that have been found for this sample. It is apparent that after the first transformation, the parameter $\theta_1$ still exhibits residual kurtosis.}
 \label{fig:wall_twopass_protocol}
\end{figure*}
\begin{figure*}
  \centering
  \begin{minipage}{0.49\textwidth}
   \centering
  \includegraphics[width=\linewidth]{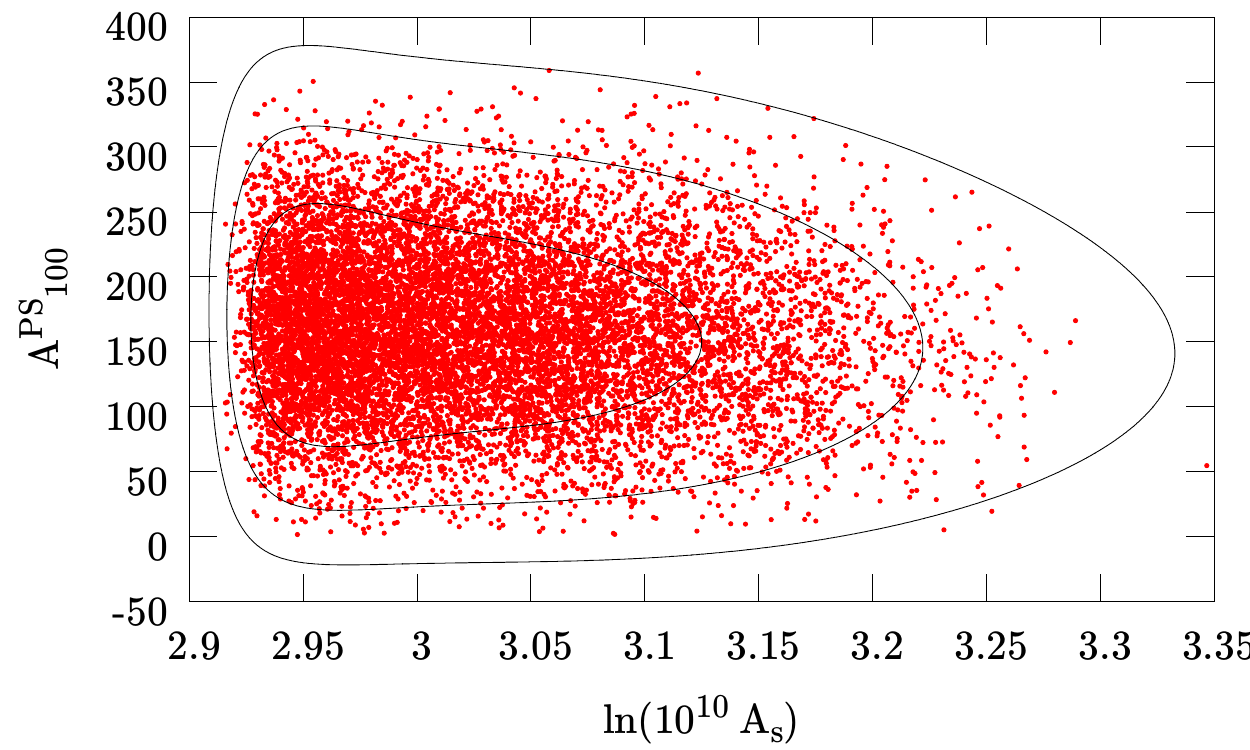}
  \end{minipage}
  \begin{minipage}{0.49\textwidth}
   \centering
   \includegraphics[width=\linewidth]{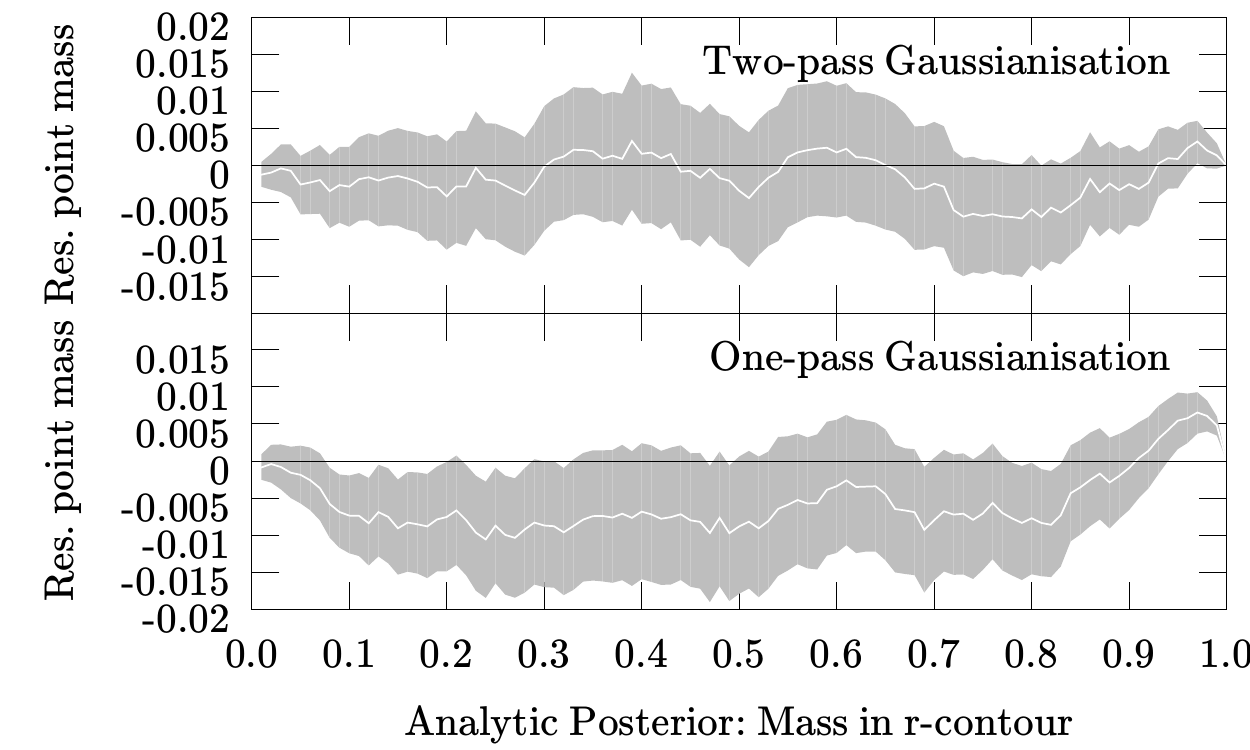}
  \end{minipage}
   \caption{One-pass Gaussianisation of a 2D marginal {\it Planck} posterior via ABC transformation. {\it Left}: original sample and contours of the analytic posterior. {\it Top right}: CC plot for the two-pass Gaussianisation in Fig.~\ref{fig:wall_twopass_protocol}. {\it Bottom right}: CC plot for the one-pass Gaussianisation (see left panel), showing deviations of the cross-contour masses.}
   \label{fig:wall_onepass}
\end{figure*}
The chains, as presented, are not decorrelated, so we thin them by using every 20th sample. We employ the \verb|`..._planck_lowl_...'| chains, which use only the temperature-temperature correlations. The plots in this section are created using the seven-parameter model including $\Omega_K$; the sample contains 11,546 points after thinning.\\
All these Markov chains assume uniform proper prior densities (i.e. being supported on compact rectangular boxes) and list the log-likelihood for every point (for further details, see \citealp{Planck13}).\\
We show several 2D marginalised posterior samples exhibiting different non-Gaussian features, and how well they are reproduced by the analytic posterior (Eq.~\ref{eq:ap}), such as triangular shapes (see Fig.~\ref{fig:triangle_onepass}), pronounced non-linear degeneracies (see Fig.~\ref{fig:banana_twopass_protocol}), and sharp boundaries (`walls') arising from MP space boundaries (see Fig.~\ref{fig:wall_twopass_protocol}).\\ Figure~\ref{fig:banana_twopass_protocol} demonstrates the usefulness of the intermediate PCA in between the ABC transformations: the first transformation has straightened out the curved shape of the maximum, but the distribution still appears skewed towards the upper left direction (see top right panel). This is remedied by reshaping, PCA and another ABC transformation (bottom right panel) -- the second Gaussianising transformation having only little effect compared to the PCA.\\
The Gaussianisation of the distribution in Fig.~\ref{fig:wall_twopass_protocol} (top left) shows how two concatenated transformations can be more powerful than a single one. The once-transformed sample still exhibits negative excess kurtosis, which is removed by the second transformation (bottom left to bottom right panel). \\
Further, we compare the CC plots of this two-pass transformation and the one-pass transformation in Fig.~\ref{fig:wall_onepass}, which also shows the resulting contours (left panel). The associated one-pass CC plot (bottom right) shows a significant deficit of point sample mass compared to the analytic posterior mass, for the posterior contours between $\sim0.1$ and $\sim0.3$, as well as for $\sim0.8$, and between $\sim0.95$ and 1. The latter is visible between the 2$\sigma$- and 3$\sigma$-contours close to the wall-like constraint at $\ln(10^{10}A_S)\simeq2.92$. By contrast, the CC plot for the two-pass transformation (Fig.~\ref{fig:wall_onepass}, top right) demonstrates good agreement between the contours of analytic posterior and point samples.\\
To demonstrate the algorithm working on a high-dimensional example, we Gaussianise a seven-dimensional {\it Planck} MCMC sample with an ABC transformation. In order to visualise the result, we show all one-dimensional and two-dimensional marginal distributions of the point sample and the full analytic posterior density (see Fig.~\ref{fig:stairs}). We employ one-pass transformations, because, as discussed in Sect.~\ref{ssec:multipass}, the marginalisation of the analytic posterior from 7D down to 2D or 1D would not be possible without explicit integration or sampling, had we chosen to use the two-pass protocol.

\section{Application: Fast evidence computation}
\label{sec:evidence}
\begin{figure}
 \centering
   \includegraphics[width=\linewidth]{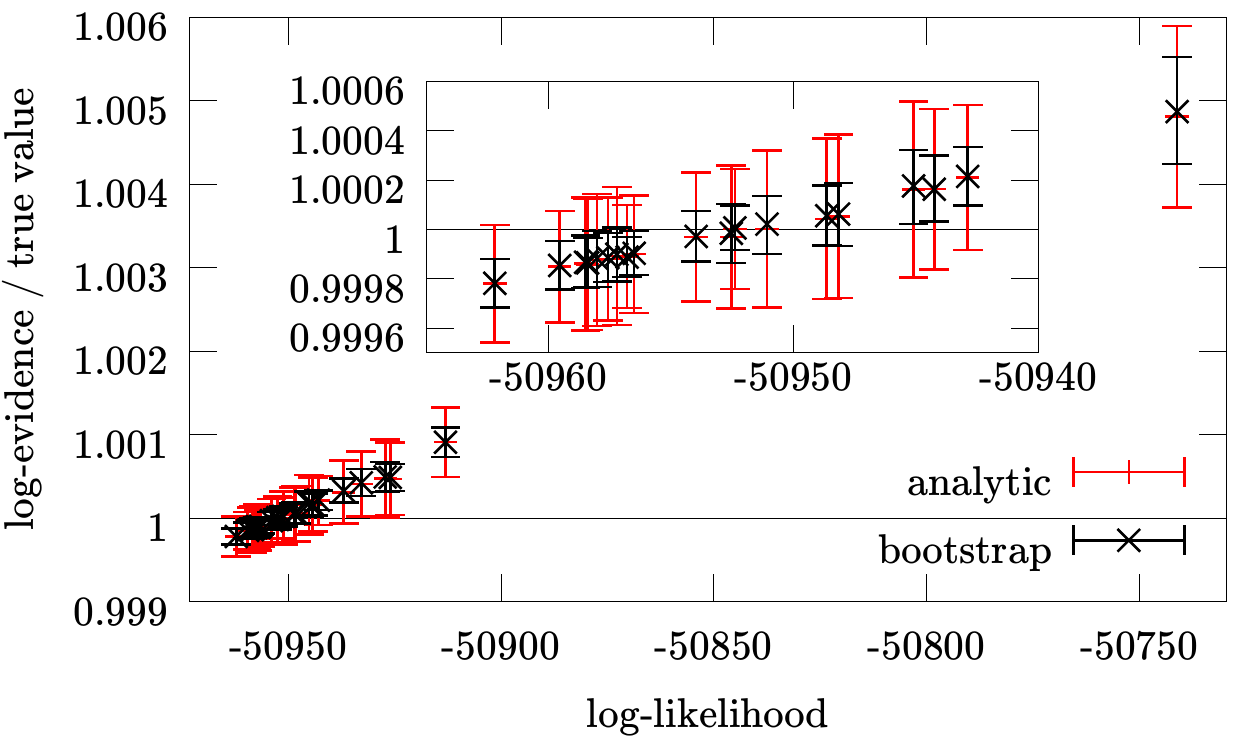}
   \caption{Ratio of log-evidence values for mock data set drawn from a 10-dimensional log-normal distribution, computed after 24 independent Gaussianisation runs to the true value, versus the negative log-likelihood for each run (including the penalty term). Values and error bars shown are from analytic regression, as detailed in App.~\ref{app:evidenceregression} (red), and the means and one-sigma confidence intervals from the bootstrapped distributions (black). The black line indicates the true value. The inset is a zoomed-in version of the lower left corner.}
 \label{fig:evidence}
\end{figure}

To decide which of two models $\M_1$ and $\M_2$, each with their associated MP space, is more predictive on a common set of data $\D$, the essential quantity to compute is the model evidence $E_i=\P(\D|\M_i)$, see \cite{jaynes, mackay, bayesfactors, skilling}. The ratio of the evidences (called Bayes factor) is then used for updating the prior model odds $\P(\M_1):\P(\M_2)$ to posterior model odds $\P(\M_1|\D):\P(\M_2|\D)$ in the Bayesian sense:
\begin{equation}
\frac{\P(\M_1|\D)}{\P(\M_2|\D)}=\frac{\P(\D|\M_1)}{\P(\D|\M_2)}\,\frac{\P(\M_1)}{\P(\M_2)}.
\end{equation}
The evidence itself, for each model, can be computed via
\begin{equation}
E_i=\P(\D|\M_i)=\int \d\bmath{X}\,\P(\D|\bmath{X}, \M_i)\P(\bmath{X}|\M_i)
\end{equation}
i.e. via integration of the (unnormalised) posterior density $\Pi(\bmath{X})=\P(\D|\bmath{X}, \M_i)\P(\bmath{X}|\M_i)$ over the respective parameter space of model $\M_i$ -- hence the term `marginal likelihood' for $E$. If $\Pi$ takes a form with non-Gaussian features, and if the model parameter space is high-dimensional, this integral itself is often difficult to calculate.\newline
However, with a bijective transformation $T:\bmath{X}\mapsto\bmath{Y}$ that Gaussianises $\widetilde{\Pi}(\bmath{Y})=\Pi[T^{-1}(\bmath{Y})]\;|\d\bmath{X}/\d\bmath{Y}|$, we can compute the evidence integral analytically. If $\widetilde{\Pi}$ has the shape of an (unnormalised) multivariate Gaussian
\begin{equation}
 \widetilde{\Pi}(\bmath{Y})=\widehat{\Pi}\;\exp\left[-\frac{1}{2}(\bmath{Y}-\tilde\bmu)^T\bmath{\tilde \Sigma}^{-1}(\bmath{Y}-\tilde{\bmu})\right]
\end{equation}
with means $\tilde{\bmu}$, covariance matrix $\bmath{\tilde{\Sigma}}$ and maximum $\widehat{\Pi}$, the log-evidence reads
\begin{equation}
\ln E = \ln\widehat{\Pi} + \frac{1}{2}\ln\det\bmath{\tilde{\Sigma}} + \frac{d}{2}\ln(2\pi).\label{eq:logevidence}
\end{equation}
Similar expressions for Gaussian posterior densities can be found in \cite{kitching}. To estimate $\widehat{\Pi}$, we need the absolute normalisation of $\widetilde\Pi$; hence this method can only be applied to samples which provide the values for $\Pi$ (possibly also in the form of log-likelihood and log-prior). From these, we compute the values of $\ln\widetilde{\Pi}(\bmath{Y})$ on the optimally-Gaussianised sample by adding the logarithm of the transformation Jacobian, and then fit the parameters $\tilde{\bmu}$, $\tilde{\bmath\Sigma}$, and $\widehat{\Pi}$ of the Gaussian via least-squares regression. This can be performed analytically, and even be used to compute an error bar on the value of $\ln E$ -- see Appendix~\ref{app:evidenceregression} for details.\newline
If the prior distribution for one MP, and hence the posterior, is supported only on a finite interval, the same will hold true for the transformed MP if we restrict ourselves to one-pass transformations. If the sample size is large enough to properly represent the cutoff, the Gaussianisation transformation will alleviate this feature, but may not fully remove it. Assuming the marginal distribution to be Gaussian, when in reality we may deal with a truncated Gaussian, will lead to a systematic error in the evidence, so it is advantageous to remove these features before starting the search for optimally Gaussianising TPs. Appendix~\ref{app:unboxing} details `unboxing transformations', which redefine the MPs, mapping a finite open interval to the entire real line. In fact, it is also possible to use them for posterior density reconstruction, before Step 1 in Sect.~\ref{ssec:multipass}.\newline
To demonstrate this idea, we compute the evidence integral first on a mock data set, and subsequently on real data from cosmology. For the former, we draw a random sample of length 10,000 from a ten-dimensional log-normal probability distribution, and assign to each point the value of the probability density function, multiplied with a factor of $E=\exp(5)$. All weights are set to unity. This mock sample is subjected first to the Gaussianisation procedure with one-pass ABC transformations (no unboxing), and then to the regression outlined in App.~\ref{app:evidenceregression} to retrieve the best-fit estimate for $\ln E$ and its error bar. To verify these, we determine the distribution of the estimator in Eq.~(\ref{eq:logevidence}) by producing 1,000 bootstrap samples from the transformed sample, and computing $\ln E$ on each of them, together with its one-sigma confidence intervals. To increase the reliability of the optimally Gaussianising transformation, 24 independent Gaussianisation runs are started with randomised initial conditions; Fig.~\ref{fig:evidence} shows the results for these. Relative to the ``true'' value of 5, these agree to sub-percent accuracy. It is also noteworthy that a lower local maximum of the likelihood, i.e. one further to the right, will depart more from the true value. Hence, a location in TP space that is close to the exact optimum will yield a biased value for the log-evidence. This indicates that the evidence is a sensitive indicator for departures from Gaussianity. Note that a log-normal distribution can be precisely Gaussianised with Box-Cox transformations -- and thus also by ABC transformations, which are a superset of these. Further, it is noteworthy that our analytic procedure yields error bars of the right magnitude, yet somewhat more conservative, compared to their bootstrapped counterparts. This discrepancy arises because the bootstrapped distributions for $\ln\det\bmath{\tilde{\Sigma}}$ and $\ln\widehat{\Pi}$ deviate slightly from Gaussianity, whereas the analytic error bars assume Gaussian error propagation (see App.~\ref{app:evidenceregression} for details).
\\
For a demonstration on real-world data, we Gaussianise the joint posterior distribution of MPs of data from weak lensing and baryon acoustic oscillations. The weak lensing data set is the 2D cosmic shear data taken by the Canada-France-Hawaii Telescope Lensing Survey (CFHTLenS; see \citealt{heymans12, kilbinger13}). The CFHTLenS survey analysis combined weak lensing data processing with THELI (\citealt{erben}), shear measurement with lensfit (\citealt{miller}), and photometric redshift measurement with PSF-matched photometry (\citealt{hildebr}). A full systematic error analysis of the shear measurements in combination with the photometric redshifts is presented in \cite{heymans12}, with additional error analyses of the photometric redshift measurements presented in \cite{benj}.\\
The BAO data set is the Data Release 9 (DR9) CMASS sample from the Baryon Oscillation Spectroscopic Survey (BOSS), which is part of the Sloan Digital Sky Survey III (SDSS-III) -- see \cite{anderson}. This contains 264,283 massive galaxies in a redshift range $0.43<z<0.7$, whose correlation function and power spectrum both exhibit the features of baryon acoustic oscillations. The quantity $d(z)=r_S(z_d) / D_V (z)$, i.e. the ratio of the comoving sound horizon $r_S$ at the baryon drag epoch $z_d$ and the spherically volume-averaged distance $D_V(z)$, is a probe of the underlying cosmological parameters -- see \cite{percival} for details.\\
To draw samples from the posterior distribution, we use the {\tt CosmoPMC} software package\footnote{See {\tt http://www2.iap.fr/users/kilbinge/CosmoPMC/.}}, which uses Population Monte Carlo (PMC), an algorithm to approximate the target distribution by a Gaussian mixture model. We compare three cosmological models: standard flat $\Lambda$CDM, curved $\Lambda$CDM, flat $w$CDM, and curved $w$CDM. The first has a four-dimensional parameter space spanned by matter density $\Omega_m$, power spectrum normalisation $\sigma_8$, baryon density $\Omega_b$, and the normalised Hubble parameter $h_{100}$ -- all other parameters are set to their best fit values for flat $\Lambda$CDM, see \cite{Planck15}. The latter two contain a fifth model variable each -- curvature parameter $\Omega_K$ and constant dark energy equation-of-state parameter $w$, respectively. For all of these parameters, flat proper priors were chosen.The baseline model -- flat $\Lambda$CDM is always referred to as model 1, whereas model 2 is one of the two extensions. As a byproduct of the sampling process, PMC provides the model evidence for the data set used -- see \cite{PMCevidence} for further details. \\
\begin{table*}
\centering
\begin{tabular}{l|c c c}
 \hline
 & flat $\Lambda$CDM & curved $\Lambda$CDM & flat $w$CDM \\ 
 \hline
 dimension & 4 & 5 & 5 \\ \\
 $\ln E$ (G) & 486.96 $\pm$ 0.01 & 485.79 $\pm$ 0.03& 486.09 $\pm$ 0.05\\
 $\ln E$ (PMC) & 487.02 $\pm$ 0.03 & 485.84 $\pm$ 0.01 & 486.00 $\pm$ 0.04\\ \\
 $\ln B_{12}$ (G) & n.a. & 1.17 $\pm$ 0.04 & 0.87 $\pm$ 0.05 \\
 $\ln B_{12}$ (SDDR) & n.a. & 1.23 $\pm$ 0.04 & 0.93 $\pm$ 0.06 \\
 $\ln B_{12}$ (PMC) & n.a. & 1.19 $\pm$ 0.04 &  1.03 $\pm$ 0.05
\end{tabular}
\caption{Values for evidence and Bayes factor for CFHTLenS+BOSS data set in three cosmological models, as computed with Box-Cox Gaussianisation (G) of weighted samples with 10,000 points each. For comparison: evidence value $\ln E$ and Bayes factor $\ln B_{12} = \ln E_\mathrm{base} - \ln E_\mathrm{extension}$ from Population Monte Carlo (PMC), and Bayes factor from Savage-Dickey density ratio (SDDR).}
\label{tab:logE}
\end{table*}
\noindent In the special situation where one model is nested inside the other, the evidence ratio $B_{12}=E_1/E_2$ can be computed via the Savage--Dickey density ratio (SDDR) -- see \cite{sddr, drsddr}, and citations therein. Under mild conditions on prior and posterior densities for the full model $\M_2$ and the submodel $\M_1$, the ratio can be derived to be 
\begin{equation}
 B_{12}=\frac{\P(\psi=\psi_\mathrm{sub}|\D, \M_2)}{\P(\psi=\psi_\mathrm{sub}|\M_2)},
\end{equation}
where $\psi$ denotes the extra parameter (or parameters) contained in $\M_2$ but not in $\M_1$, $\psi_\mathrm{sub}$ is the value of $\psi$ that specifies the submodel $\M_1$, and $\P(\psi|\D, \M_2)$ and $\P(\psi|\M_2)$ are posterior and prior densities of the full model, marginalised over all model parameters but $\psi$ (see \citealp{drsddr} for details).\\
We find that the log-evidence values computed by {\tt CosmoPMC} need to be offset by a factor of $n-1$ times the log-prior density, where $n$ is the number of data sets used. This is due to a non-standard interpretation of the prior density within {\tt CosmoPMC}. Throughout this work, we apply this correction to the log-evidence values produced by {\tt CosmoPMC} as well as the log-posterior values extracted from the {\tt CosmoPMC} output. We follow the practice of (\citealp{PMCevidence, kilbinger13}) of accepting a {\tt CosmoPMC} run as soon as the built-in convergence diagnostic, called perplexity, exceeds a value of  $p>0.7$. Sampling to even higher values for the perplexity, up to $p\sim0.95$, still changes the {\tt CosmoPMC} value for $\ln E$ by as much as $\sim$ 0.1 -- this indicates a residual bias in the statistic. However, since the exact same offset has to appear in the {\tt CosmoPMC} output values for the log-evidence $\ln E$ and for the non-normalised log-posterior $\ln\Pi(\bmath X)$, it is not of relevance to demonstrating our method, so investigating its origin is beyond the scope of this work.\\ 
Table~\ref{tab:logE} shows the log-evidences for the three models, and the Bayes factors of $\Lambda$CDM compared to either of the two extended models. The numbers in the first line were computed via one-pass Gaussianisation with ABC transformations, preceded by an unboxing transformation. To estimate the scatter of the {\tt CosmoPMC} and SDDR values for $\ln E$ and $\ln B_{12}$ in the second, fourth, and fifth lines, we rerun {\tt CosmoPMC} ten times for each model, and determine the mean and average for the {\tt CosmoPMC} and SDDR estimators. Like for the log-normal sample, 24 independent Gaussianisation runs were started for each sample, and the one with the highest log-likelihood value chosen to transform the sample, which is then subjected to the analytic evidence computation procedure. The values in the first row of Table~\ref{tab:logE} are the weighted averages and standard deviations of all ten values, where the weights are determined from the analytic error bars as $w_i=\sigma_i^{-2}$ (cf. App.~\ref{app:evidenceregression}). The values show that the combined data favour $\Lambda$CDM over any of the extended models, although the evidence is not strong against either of the two.
Our values agree with the numbers of SDDR and PMC within the spread between the latter two estimators, but still small deviations remain, which are larger than the error bars quoted. These may be due to residual non-Gaussianity in the transformed samples, to which the evidence is a sensitive measure.

\section{Conclusions and outlook}
\label{sec:conclusio}
\begin{description}
\item We have discussed how to transform a posterior probability density approximately into a multivariate Gaussian, and various applications thereof:
\begin{enumerate}
\renewcommand{\theenumi}{(\arabic{enumi})}
\item From the parameters of the Gaussianised distribution and those of the transformation, we can reconstruct an analytic expression for the original posterior probability density, given a point sample drawn from it. This facilitates the combination of different data sets to obtain the joint posterior density.
\item Further, this analytic posterior can be used to display contours of the density in question or its marginals, without the need for density estimates or smoothing procedures. Also, in reproducing the contours of the probability density reliably, it outperforms kernel density estimates.
\item We suggest that, instead of distributing lengthy point samples in the form of a Markov chain, to use a Gaussianising transformation to disseminate a posterior density. Only the transformation parameters and the first and second moments of the resulting Gaussian are needed to reproduce the posterior density in its functional form; hence we can achieve substantial data compression.
\end{enumerate}

\item We have demonstrated this algorithm with our implementation in {\tt C} (code on request), which employs Markov Chain Monte Carlo (MCMC) samples from {\it Planck} data. We used Box--Cox and Arcsinh--Box--Cox (ABC) transformations to Gaussianise various marginal distributions with distinctive non-Gaussian features, and showed the resulting contours.\\

\item One distinctive application of Gaussianising transformations, which we discuss and demonstrate here, is a novel method to compute the model evidence of a posterior distribution, given a point sample from it. We have tested this method on cosmological data from lensing and baryon acoustic oscillations, for different cosmological models, and find slight preference for $\Lambda$CDM. Compared to the numerical results from Population Monte Carlo (PMC) and the Savage-Dickey Density Ratio (SDDR), our new method of computing the evidence agrees well within the spread of the other two.\\

\item We have introduced the cross-contour (CC) plot as a tool to decide whether one probability density reproduces the contours of another, or if they do not, to detect where they deviate. \\

\end{description}
There are several possible extensions of our method, and directions to advance its scope: 
\begin{description}
\item To optimise the Gaussianisation algorithm for speed and/or accuracy, it is possible to replace the Nelder--Mead minimum finder with other, more sophisticated algorithms, such as simulated annealing (\citealt{simann}), or BOBYQA (\citealt{bobyqa}).\\

\item It is possible to engage new families of transformations, designed to cure a wider spectrum of non-Gaussian features that a multivariate probability density may possess -- in our implementation, new families can easily be included.\\

\item Gaussianisation may be employed for fast sampling from a non-Gaussian probability density, in case that the Gaussianising parameters are either known exactly or to sufficient accuracy. Afterwards, it is possible to quickly draw a point sample from a multivariate Gaussian distribution, and transform this sample with the inverse map.\\

\item To improve the accuracy of the evidence computation, it is possible to replace the log-likelihood of Eq.~(\ref{eq:like}) with another loss function, which penalises deviations from Gaussianity in a sharper manner.\\

\item So far, we have been working with unimodal probability densities. We require the transformations to be bijective, hence we cannot map a multimodal distribution into a unimodal Gaussian. However, we may be able to transform such a density into a mixture of (possibly overlapping) Gaussians, where we now have to estimate the weight factor for each constituent from the transformed sample, in addition to each $\tilde\bmu$ and $\bmath{\tilde{\Sigma}}$. The requisite number of components can possibly be determined with standard clustering algorithms.
\end{description}

\section*{Acknowledgements}
RLS is grateful to Martin Kilbinger, Karim Benabed, Catherine Heymans and Stephen Feeney for helpful discussions, and has been supported by the Perren Fund in Astronomy, and by IMPACT (UCL MAPS faculty); BJ acknowledges support by an STFC Ernest Rutherford Fellowship, grant reference ST/J004421/1; HVP was supported by the European Research Council under the European Community's Seventh Framework Programme (FP7/2007-2013) / ERC grant agreement no.306478-CosmicDawn. This work is based on observations obtained with Planck ({\tt http://www.esa.int/Planck}), an ESA science mission with instruments and contributions directly funded by ESA Member States, NASA, and Canada. In part, this work was supported by National Science Foundation Grant No. PHYS-1066293 and the hospitality of the Aspen Center for Physics. 

\appendix

\section{Analytic evidence computation}
\label{app:evidenceregression}
We outline how the computation of the log-evidence (see Eq.~\ref{eq:logevidence}) can be performed analytically, i.e. without numerical optimisation. Our data consist of a Gaussianised weighted sample of $\mathcal{N}$ points in $\mathbfss{R}^d$, $\{(\bmath{Y}^a, w^a)\}_{a=1}^\mathcal{N}$ and the values of the transformed log-posterior on each of these points, $\{\ell^a\}_{a=1}^\mathcal{N}$. To fit a multivariate unnormalised Gaussian 
\begin{equation}
 \widetilde{\Pi}(\bmath{Y})=\widehat{\Pi}\exp\left[-\frac{1}{2}(\bmath{Y}-\tilde\bmu)^T\bmath{\tilde \Sigma}^{-1}(\bmath{Y}-\tilde{\bmu})\right]\label{eq:unnormgauss}
\end{equation}
through the values of  $\{\exp(\ell^a)\}_{a=1}^\mathcal{N}$, we use the regression model
\begin{equation}
 \ell_{\bmath{A,B},C}^\mathrm{model}(\bmath Y)=\bmath{Y}^T\bmath{A\;Y}+\bmath{B}^T\bmath{Y}+C,
\end{equation}
which is linear in each of the $d(d+3)/2+1$ regression parameters: the upper-diagonal components of the symmetric matrix $\bmath{A}$, the components of vector $\bmath{B}$, and the scalar $C$. Assuming independence and homoscedasticity, we arrive at our quantity to minimise,
\begin{equation}
\chi^2(\bmath{A}, \bmath{B}, C)=\sum\limits_{a=1}^{\mathcal{N}}w^a\left[\ell^\mathrm{model}_{\bmath{A}, \bmath{B}, C}(\bmath{Y}^a)-\ell^a\right]^2,
\end{equation}
which is quadratic in every regression parameter. Thus, we can write all normal equations of the regression problem,
\begin{equation}
\frac{\d\chi^2}{\d\vartheta}\stackrel{!}{=}0$,\quad $\vartheta\in\{A_{ij}, B_k,C\}
\end{equation} as a $[d(d+3)/2+1]$-dimensional linear inhomogeneous vector equation, and solve via singular value decomposition. From the resulting values of $(\bmath{A}, \bmath{B}, C)$, the parameters of the multivariate Gaussian (Eq.~\ref{eq:unnormgauss}) can readily be computed as
\begin{eqnarray}
 \bmath{\tilde{\Sigma}}&=&-\frac{1}{2}\bmath{A}^{-1};\\
 \tilde{\bmu}&=&-\frac{1}{2}\bmath{A}^{-1}\bmath{B};\\
 \ln\widehat{\Pi}&=&C-\frac{1}{4}\bmath{B}^T\bmath{A}^{-1}\bmath{B}.
\end{eqnarray}
Furthermore, we can use the analytic regression procedure to find error bars on these estimators, and thus on $\ln E$. To this end, we can analytically find the covariance matrix $\bmath{Cov}$ of all parameters $\{A_{ij}, B_i, C\}$ from the form of $\chi^2$ and the transformed data set $\{(\bmath{Y}^a, w^a, \ell^a)\}_{a=1}^\mathcal{N}$, and then approximate the variances of $\ln\widehat{\Pi}$ and of $\ln\det\bmath{\tilde\Sigma}$ by standard Gaussian error propagation. In particular:
\begin{eqnarray}
\V{\ln E} &=& \V{\ln\widehat{\Pi}}+\frac{1}{4}\V{\ln\det\bmath{\tilde{\Sigma}}} \\
&\simeq&\bmath{\Xi}^T\bmath{Cov}\,\bmath{\Xi}+\frac{1}{4}\bmath{\Upsilon}^T\bmath{Cov}\bmath{\Upsilon}\label{eq:variance}
\end{eqnarray}
with
\begin{equation}
\bmath{\Xi}=\left(\frac{\partial \ln\widehat{\Pi}}{\partial\vartheta_i}\right)=\left(
\begin{array}{cr}
\vdots&\\
\frac{1}{4}\bmath{B}'_m\bmath{B}'_n(2-\delta_{mn}) & (m=1...d,\\
\vdots&n=m...d)\\
\hline\\
\vdots&\\
-\frac{1}{2}\bmath{B}'_k&(k=1...d)\\
\vdots&\\
\hline\\
1&
\end{array}\right)
\end{equation}
and
\begin{eqnarray}
\bmath{\Upsilon}&=&\left(\frac{\partial \ln\det\bmath{\tilde{\Sigma}}}{\partial\vartheta_i}\right)\nonumber\\
&=&\left(
\begin{array}{cr}
\vdots&\\
\frac{1}{4}(\bmath{A}^{-1})_{mn}(2-\delta_{mn}) & (m=1...d,\\
\vdots&n=m...d)\\
\hline\\
\vdots&\\
0&\\
\vdots&\\
\hline\\
0&
\end{array}\right),
\end{eqnarray}
where $\bmath{B}'=\bmath{A}^{-1}\bmath{B}$.\newline

\section{Unboxing Transformations}
\label{app:unboxing}
A single MP $Z$, which is assumed to be constrained to an open interval $(a, b)$, is redefined via the unboxing transformation $U_{(a,b)}:(a,b)\rightarrow\mathbfss{R}$
\begin{equation}
X=U_{(a,b)}(Z)=\frac{a+b}{2}+\frac{b-a}{\sqrt{2\pi}}\Phi^{-1}\left(\frac{Z-a}{b-a}\right),
\end{equation}
where $\Phi^{-1}$ denotes the inverse of the cumulative distribution function of the Normal distribution:
\begin{equation}
\Phi(x)=\int_{-\infty}^{x}dy\frac{1}{\sqrt{2\pi}}\exp\left(-\frac{y^2}{2}\right).
\end{equation}
\label{lastpage}
$U_{(a,b)}$, thus designed, has the following properties: it is bijective and smooth; the limits are $\lim_{Z\rightarrow a}U_{(a,b)}(Z)=-\infty$; $\lim_{Z\rightarrow b}U_{(a,b)}(Z)=+\infty$. Further, the midpoint of the interval $m=\frac{1}{2}(a+b)$ is fixed: $U_{(a,b)}(m)=m$, $U'_{(a,b)}(m)=1$. Sending the interval boundaries to infinity simultaneously will result in the identity transformation
\begin{equation}
\lim_{a\rightarrow-\infty\atop b\rightarrow+\infty}U_{(a,b)}(Z)=Z.
\end{equation}
This is a generalisation of the widely used probit transformation, which maps the unit interval onto the real numbers as $p\mapsto\Phi^{-1}(p)$. Our modified probit has one huge advantage for the subsequent search for a Gaussianising transformation: If $Z$, as a random variable, is uniformly distributed on $(a,b)$, then $X$ is normally distributed with mean $m$ and spread $(b-a)/\sqrt{2\pi}$. \newline
In statistics, a frequently-used alternative to probit is the logit map $p\mapsto \log(p/1-p)$. For our purposes, however, the probit is preferable, since a similarly rescaled version of this logit-transformation would yield a distribution with excess kurtosis, instead of a Gaussian.\newline
For a $d$-dimensional vector of MPs $\bmath{Z}=(Z_1,\ldots, Z_d)$, constrained to intervals $(a_i,b_i)\ni Z_i$, we unbox each dimension separately, with the appropriate boundaries:
\begin{equation}
\bmath{Z}\mapsto\bmath{X}=\left[U_{(a_1,b_1)}(Z_1),\ldots, U_{(a_d,b_d)}(Z_d)\right].
\end{equation}
Before starting the search for the Gaussianisation parameters, every point in the original sample is mapped through this transformation.
\end{document}